\documentclass[aps,prd,reprint,superscriptaddress,nofootinbib]{revtex4-2}

\usepackage{graphicx}
\usepackage{dcolumn}
\usepackage{bm}
\usepackage{hyperref} 
\usepackage[mathlines]{lineno}
\usepackage{amsmath}
\usepackage{dsfont}
\usepackage{float}
\usepackage{mathrsfs}
\usepackage{soul}
\usepackage[dvipsnames]{xcolor}
\usepackage{orcidlink}

\newcommand{\vR}[0]{\varrho}

\begin{document}
\date{\today}

\title{Evolution and the quasistationary state of collective fast neutrino flavor conversion in three-dimensions without axisymmetry} 
%
\author{Manu George\orcidlink{0000-0002-8974-5986}}
\email{kuttan.mgc@gmail.com}
\affiliation{Institute of Physics, Academia Sinica, Taipei 115201, Taiwan}
\author{Zewei Xiong\orcidlink{0000-0002-2385-6771}}
\email{z.xiong@gsi.de}
\affiliation{GSI Helmholtzzentrum {f\"ur} Schwerionenforschung, Planckstra{\ss}e 1, 64291 Darmstadt, Germany}
\author{Meng-Ru Wu\orcidlink{0000-0003-4960-8706}}
\email{mwu@gate.sinica.edu.tw}
\affiliation{Institute of Physics, Academia Sinica, Taipei 115201, Taiwan}
\affiliation{Institute of Astronomy and Astrophysics, Academia Sinica, Taipei 106319, Taiwan}
\affiliation{Physics Division, National Center for Theoretical Sciences, Taipei 106319, Taiwan}
\author{Chun-Yu Lin\orcidlink{0000-0002-7489-7418}}
\email{lincy@nchc.org.tw}
\affiliation{National Center for High-performance Computing, Hsinchu 30076, Taiwan}

\begin{abstract}
We investigate in this work the evolution of the collective fast neutrino flavor conversion (FFC) in a three-dimensional (3D) cubic box with periodic boundary condition for three different neutrino angular distributions that are axially asymmetric. 
We find that the system evolves toward a quasistationary state where the angular distribution of the spatially averaged neutrino electron-minus-muon lepton number (ELN)  does not contain any crossings. 
In the quasistationary state, near flavor equilibration is achieved in one angular domain enclosed by the initial ELN angular crossing contour, similar to the conclusion derived based on simplified one-dimensional (1D) system with axially symmetric neutrino angular distributions.  
We have also performed additional simulations in coordinates where the initial first ELN angular moment has only one nonvanishing spatial component by using the original axially asymmetric ELN angular distributions as well as the corresponding axisymmetric ELN distributions and find interesting similarity between these two sets. 
Finally, we propose three different analytical prescriptions generalized from earlier 1D models to 3D models and evaluate their performances in predicting the post-FFC moments. 
Our findings suggest that further development of effective classical transport models in multidimensions to capture the effect of FFC is promising. 
\end{abstract}

\maketitle
\graphicspath{{./}{./figures/}{./static}}

\section{Introduction}
Neutrinos are known to have significant impacts on the dynamics and composition of astrophysical systems such as the core-collapse supernovae and the neutron star mergers.
In the central regions of these systems where the neutrino densities are large, the flavor evolution of neutrinos is dominated by the nonlinear self-coupling due to the coherent neutrino-neutrino forward scattering~\cite{pantaleone1992neutrino,sigl1993general}. 
The nonlinear nature leads to various collective phenomena of neutrino flavor oscillations (see e.g.~\cite{capozzi2022neutrino,volpe2024neutrinos,fischer2024neutrinos} for recent reviews)
and can potentially affect our understanding of these astrophysical events \cite{stapleford2020coupling,xiong2020potential,li2021neutrino,just2022fast,fernandez2022fast,fujimoto2022explosive,nagakura2023roles,grohs2023neutrino,ehring2023fast,ehring2023fast1,xiong2024fast}.
Among these, the likely occurrence of fast flavor conversion (FFC)~\cite{sawyer2009multiangle,sawyer2016neutrino} due to the presence of the angular crossing in the neutrino electron-minus-muon lepton number (ELN) in the supernova core or in the merger remnant has been identified \cite{abbar2019occurrence,delfanazari2019linear,delfanazari2020fast,nagakura2019fastpairwise,abbar2020fast,glas2020fast,nagakura2021where,abbar2021characteristics,harada2022prospects,wu2017fast,wu2017imprints,george2020fast,george2020fast,li2021neutrino,just2022fast,richers2022evaluating,fernandez2022fast,mukhopadhyay2024time}. 
This has triggered a tremendous number of studies over the last decade on this topic; see, e.g.,~\cite{tamborra2021new,richers2022fast} for earlier reviews. 

Since FFC typically develops within physical scales of subnanoseconds and subcentimeters, much smaller than the hydrodynamical or interaction time and length scales in supernovae or neutron star mergers, one strategy to study FFC is to numerically solve the neutrino quantum kinetic equation ($\nu$QKE) \cite{sigl1993general,vlasenko2014neutrino,volpe2015neutrino} in a small local volume, within which the system is assumed to be nearly homogeneous and collisions of neutrinos may be neglected. 
Taking the periodic boundary condition, recent works that assume translation symmetry in two spatial dimensions and axisymmetry in the neutrino angular distributions in local one-dimensional (1D) boxes suggest that FFC drives the system to a quasistationary state where near flavor equilibration is achieved on one side of the ELN crossing when coarse-grained over the volume of the box  
\cite{bhattacharyya2020latetime,bhattacharyya2021fast,wu2021collective,richers2021particleincell,grohs2023neutrino,richers2022code,gfiorillo2024fast,cornelius2024perturbing,azari2024systematic}. 
On the other side of the ELN crossing, the corresponding coarse-grained properties can be described by simple formulas subject to the conservation of the total ELN \cite{zaizen2023simple,xiong2023evaluating}. 
Based on these results, Reference~\cite{xiong2024robust} recently showed that it is possible to effectively include FFC in the classical transport model of neutrinos by applying the quasistationary state prescription obtained from local simulations to locations where ELN crossings are found in global transport simulations under spherically symmetric and static supernova background profiles. 
Different methods addressing the global $\nu$QKE simulations including FFC have also been attempted~\cite{capozzi2019collisional,padilla2021multidimensional,nagakura2022timedependent,xiong2023evolution,nagakura2023basic,nagakura2023roles,nagakura2023global,xiong2024fast,shalgar2023neutrino,shalgar2023neutrino1,cornelius2024perturbing,cornelius2024neutrino} or proposed~\cite{johns2023thermodynamics,nagakura2024bgk,johns2024subgrid}. 
Efforts on identifying fast flavor instabilities \cite{abbar2021characteristics,johns2021fast,nagakura2021new,richers2022evaluating,abbar2023applications,abbar2024detecting} and predicting the post-FFC angular moments \cite{abbar2024physicsinformed,abbar2024application} based on limited available information of the neutrino angular moments as well as simulation methods that directly evolve the neutrino angular moments for FFC \cite{froustey2024neutrino,grohs2024twomoment} have been investigated. 

As simulations of FFC in multiple spatial dimensions are computationally more demanding, they have been carried out only in a handful of works \cite{richers2021neutrino,nagakura2023global,cornelius2024neutrino}. 
Ref.~\cite{richers2021neutrino} performed local simulations in two- and three-dimensional (2- and 3D) boxes with periodic boundary conditions for several cases. 
For those with zero total ELN initially, coarse-grained flavor equilibration is reached, independent of the dimensionality of the system. 
For the case with an initial nonzero total ELN that possesses axisymmetry in the ELN angular distribution, a similar coarse-grained quasistationary state is also obtained in simulations with different dimensions. 
Although these results seem to hint that the outcome of FFC in a multidimensional periodic box may be similar to that obtained in the corresponding 1D case with axisymmetry, it remains important to investigate whether such a conclusion holds for more general initial conditions\footnote{We note that, during the completion of this manuscript, Ref.~\cite{richers2024asymptoticstate} which performed similar study and analysis appeared. Our work presented here has been done independently using a different simulation code.}. 
It is worth noting that global 2D FFC simulations have been carried out to explore the impact of the initial condition \cite{cornelius2024neutrino} and the 
evolution of FFC in neutron star merger remnants~\cite{nagakura2023global}. 

In this work, we perform 3D simulations of FFC in a local box with the periodic boundary condition for three cases where the system has nonzero total ELN and does not contain any axisymmetry. 
We find that for all these cases, the ELN angular crossings are erased in the quasistationary state on a coarse-grained level. 
Near flavor equilibration is also reached in one angular domain enclosed by the initial ELN crossing contour, in agreement with what were found in earlier 1D and multidimensional studies.  
We also perform additional simulations for the same physical systems in rotated coordinates where only one spatial component of the ELN fluxes is nonvanishing, and the corresponding auxiliary simulations for cases that possess axisymmetry. 
We will show that in the rotated coordinates, the numerical results based on the auxiliary axisymmetric cases can be used to accurately describe the evolution of neutrino angular moments in the axially asymmetric cases.
These simulation outcomes allow one to generalize previously proposed analytical prescriptions  \cite{zaizen2023simple,xiong2023evaluating} to predict the post-FFC angular moment values.  

\begin{figure*}[t] 
\includegraphics[width=0.8\linewidth]{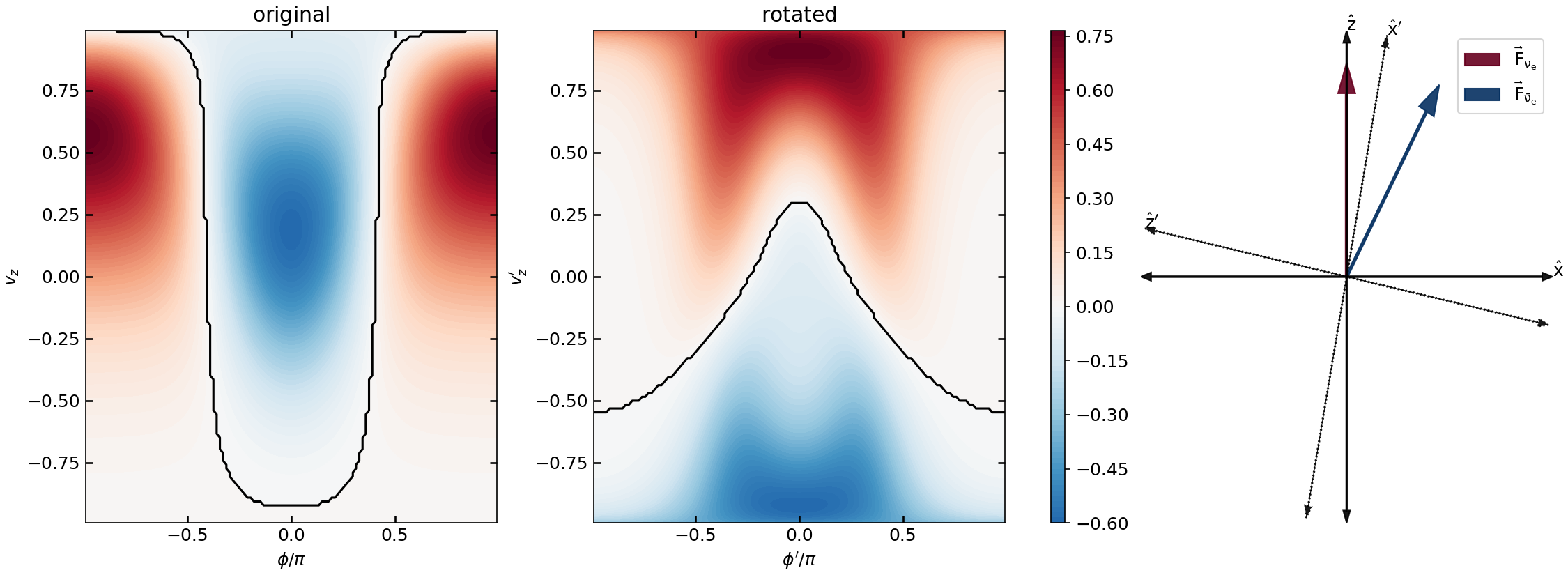}
\caption{Initial ELN angular distribution function $G_{v_z,\phi}^0$ in the original (left panel) and $G_{v'_z,\phi'}^0$ the rotated (middle panel) coordinate systems for $\theta_r=30^\circ$.  
The relation of the two coordinate systems is shown in the right panel. 
}
\label{fig:roteln30}
\end{figure*}

This paper is organized as follows. 
In Sec.~\ref{sec:setup}, we describe the setup of our model, including the equation of motion, the definition of the coordinate systems, and the numerical scheme used for the simulations.  
In Sec.~\ref{sec:results_numerical}, we present our simulation results in both coordinate systems and discuss the implications. 
In Sec.~\ref{sec:approx_prescript}, 
we show how to generalize various analytical prescriptions developed based on 1D box models to 3D cases to approximately predict the post-FFC angular moments. 
Conclusions and outlook are given in Sec.~\ref{sec:concluion}. 
Throughout this paper, natural units with $\hbar=c=1$ are adopted.

\section{Model Setup}\label{sec:setup}
\subsection{Neutrino flavor transport equations}\label{subsec:setup::eom}

We consider a simplified two-flavor neutrino system in a localized 3D box in which oscillations can convert the initial $\nu_e$ and $\bar\nu_e$ to the heavy lepton flavors $\nu_x$ and $\bar\nu_x$. 
As we focus on studying FFC in our simulation domain, 
we assume that the neutrino distribution functions inside the box are
homogeneous  
(before applying perturbation seeds; see below), and neglect the collisions for neutrinos.   
The neutrino vacuum mixing and the neutrino-matter forward scattering potentials can also be omitted for simplicity. 
Under these assumptions, the space-time evolution of the normalized neutrino and antineutrino densities, 
$\vR_{\bm v}(x)$ and $\bar\vR_{\bm v}(x)$, is governed by the following equations
\begin{subequations}\label{eq:eom_full}
\begin{align}
	\label{eq:eom_nu}
    v^\eta\partial_\eta\vR_{\bm v}(x) &= -i\left[H_{\bm v}(x), \vR_{\bm v}(x)\right], \\
    v^\eta\partial_\eta\bar\vR_{\bm v}(x) &= -i\left[H_{\bm v}^*(x), \bar\vR_{\bm v}(x)\right],
    \label{eq:eom_bnu} 
\end{align}
\end{subequations}
where $x^\eta=(t,\bm x)$, $v^\eta = (1, \bm v)$ with $v^\eta v_\eta = 0$. 
In Eq.~\eqref{eq:eom_nu}, 
\begin{align}
    H_{\bm v}(x) = & \mu \int d\Gamma' (1-\bm v \cdot \bm v') [g_\nu(\bm v')\vR_{\bm v'}(x) - g_{\bar\nu}(\bm v')\bar\vR_{\bm v'}^*(x)],
    \label{eq:eom_H}
\end{align}
where $\mu=\sqrt{2}G_F n_\nu$ with $n_\nu$ the neutrino number density and 
$d\Gamma=(dv_zd\phi)/(2\pi)$ with $v_z$ the $z$ component of $\bm v$ and $\phi$ the corresponding azimuthal angle on the $x$-$y$ plane.
The neutrino angular distribution function $g_{\nu}(\bm v)$ is normalized by $\int d\Gamma g_{\nu}(\bm v) = 1$, while the antineutrino one $g_{\bar\nu}(\bm v)$ satisfies 
$\int d\Gamma g_{\bar\nu}(\bm v) = n_{\bar\nu}/n_{\nu}=\alpha$, which represents the number density ratio between antineutrinos and neutrinos. 
The ELN angular distribution function $G_{\bm v}(x)$ is defined by
\begin{equation}\label{eq:GELNdef}
G_{\bm v}(x)= g_\nu(\bm v)(\vR_{ee}-\vR_{xx}) - g_{\bar\nu}(\bm v)(\bar\vR_{ee}-\bar\vR_{xx}), 
\end{equation}
where $\vR_{ee}$ and $\vR_{xx}$ are the diagonal entries of $\vR_{\bm v}(x)$ in the flavor basis, and $\bar\vR_{ee}$ and $\bar\vR_{xx}$ are the corresponding ones of $\bar\vR_{\bm v}(x)$, whose dependence on $\bm v$ and $x$ are not displayed explicitly.   

\subsection{Initial ELN distributions}\label{subsec:setup::angdistr}

For given initial angular distributions $g_{\nu}(\bm v)$ and 
$g_{\bar\nu}(\bm v)$ for $\nu_e$ and $\bar\nu_e$, the corresponding flux vectors normalized by the neutrino number density 
can be computed by
\begin{equation}
{\bm F}^0_{\nu_e(\bar\nu_e)} = \int d\Gamma\bm v g_{\nu(\bar\nu)}(\bm v).    
\end{equation}
Without loss of generality, we take a nonzero angle $\theta_r=\cos^{-1}[{\bm F}^0_{\nu_e}\cdot{\bm F}^0_{\bar\nu_e}/(|{\bm F}^0_{\nu_e}||{\bm F}^0_{\bar\nu_e}|)]$ between
${\bm F}^0_{\nu_e}$ and ${\bm F}^0_{\bar\nu_e}$.
We assume that $g_{\nu(\bar\nu)}(\bm v)$ are axisymmetric with respect to the direction of their respective flux vectors and are given by
\begin{equation}\label{eq:g_loc_f}
g_{\nu(\bar\nu)}(\bm v) \propto\exp[-(v_{F_{\nu(\bar\nu)}} - 1)^2/(2\sigma^2_{\nu(\bar\nu)})],
\end{equation}
where $v_{F_{\nu(\bar\nu)}}=\bm v\cdot{\bm F^0_{\nu(\bar\nu)}}/|\bm F^0_{\nu(\bar\nu)}|$ are the velocity projections in the directions of the flux vectors and $\sigma_{\nu(\bar\nu)}$ are width parameters that determine the degree of anisotropy of $g_{\nu(\bar\nu)}$. 

Throughout the rest of the paper, we adopt $\alpha=0.9$, $\sigma_\nu=0.6$, $\sigma_{\bar\nu}=0.5$, and take $\theta_r=30^\circ$, $45^\circ$, and $60^\circ$ to explore three different cases without axisymmetry along any directions. 
As will be further discussed below, for each case, we perform simulations in two different coordinate systems for the same $G_{\rm v}$ that is axially asymmetric. 
The first coordinate system is chosen such that $F^0_{\nu_e,x}=F^0_{\nu_e,y}=F^0_{\bar\nu_e,y}=0$, i.e., $\bm F^0_{\nu_e}$ is along the $z$-axis while $\bm F^0_{\bar\nu_e}$ lies on the $x$-$z$ plane, similar to what taken in Ref.~\cite{richers2021neutrino}, and it is denoted as ``original'' coordinates. 
For the second coordinate system (labeled with superscript $'$), it is rotated from the first one such that $F^0_{\nu_e,x'}-F^0_{\bar\nu_e,x'}=0$, i.e., the $x'$ component of the first ELN moment is initially zero and is denoted as ``rotated'' coordinates for the rest of the paper. 
The rotation angle between the two coordinates is given by 
\begin{equation}\label{eq:global_rotangle}
	\theta_{rot} = \tan^{-1} \Bigg( \frac{\sin\theta_{r}}{\frac{|\bm F^0_{\nu_e}|}{|\bm F^0_{\bar\nu_e}|} - \cos\theta_{r}}\Bigg).
\end{equation}
Figure~\ref{fig:roteln30} shows 
the relation between the two coordinates (right panel) as well as 
the initial ELN distributions $G^0_{v_z,\phi}$ in the original (left panel) and $G^0_{v'_z,\phi'}$ in the rotated coordinates (middle panel) for $\theta_r=30^\circ$ as an example, with the ELN crossing contours indicated by the black solid curves. 
Clearly, the initial ELN angular distribution is less axisymmetric in the original coordinates than in the rotated coordinates. 
We note that the ELN angular distributions constructed here have reflection symmetry with respect to $\phi\rightarrow -\phi$ (or $\phi' \rightarrow -\phi'$ in the rotated coordinates), resulting in vanishing $y$ ($y'$) components of all flux vectors. 

Besides the axially asymmetric ELN angular distributions described above, we also perform auxiliary simulations in the rotated coordinates by taking the axisymmetric angular distributions averaged over $\phi'$ from $g_{\nu(\bar\nu)}(\bm v')$
\begin{equation}
g^a_{\nu(\bar\nu)}(v'_z) =
\int \frac{d\phi'}{2\pi} g_{\nu(\bar\nu)}(\bm v').
\end{equation}
The corresponding ELN angular distributions $G^a_{v'_z}(x)$ can be evaluated in the same way as Eq.~\eqref{eq:GELNdef}. 

\subsection{Simulation setup}\label{subsec:setup.numerical}

We use the extended version of \textsc{cose}$\nu$, which adopts a grid-based method to solve Eq.~\eqref{eq:eom_full}, to perform numerical simulations in a 3D cubic box with periodic boundary conditions in all three spatial dimensions. 
The box has a volume of $L^3$ with $L=100$~$\mu^{-1}$ and is discretized by $N^3$ rectangular cell-centered grids with $N=100$.  
For the 2D phase (angular) space, we discretize $-1\leq v_z\leq 1$ and $0\leq\phi\leq 2\pi$ into $N_{v_z}\times N_\phi$ cell-centered bins with $N_{v_z}=32$ and $N_\phi=8$ (same for both the original and rotated coordinates). 
The spatial derivatives are evaluated using the finite-volume plus seventh-order accurate weighted essentially nonoscillatory scheme while the time integration is computed using the fourth-order Runge-Kutta method (see \cite{george2023cose} for details). 
We have taken a fixed time step size $\Delta t=C_{\rm CFL}\times(L/N)=0.4$~$\mu^{-1}$ with $C_{\rm CFL}=0.4$ the Courant-Friedrichs-Lewy number.   
The phase-space integration in Eq.~\eqref{eq:eom_H} is taken using the simple Riemann sum. 

\begin{figure*}[hbt!]
\includegraphics[width=\textwidth]{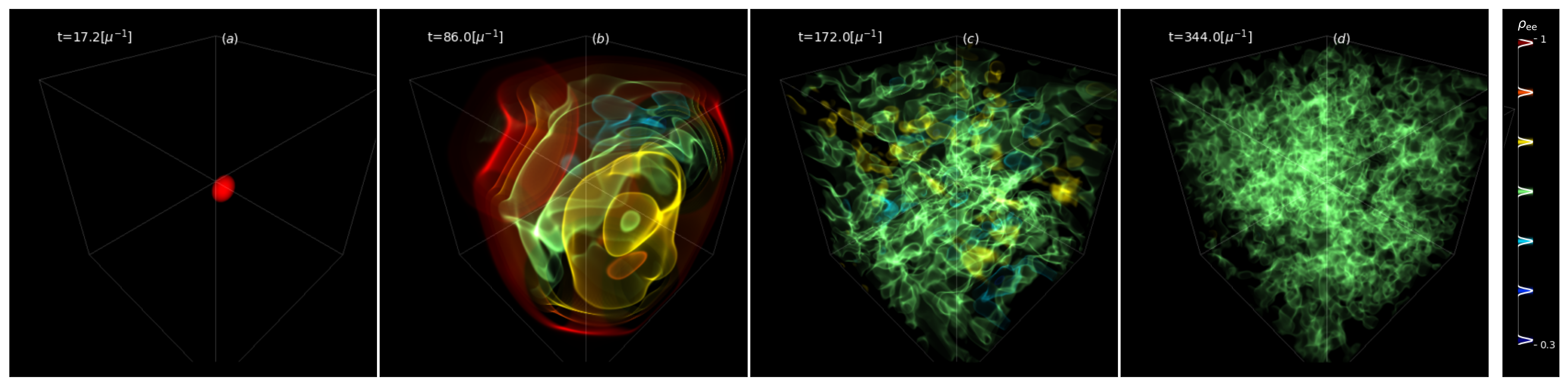}%
\caption{
Spatial distribution of of $\langle \vR_{ee}(\bm x) \rangle_\Gamma$ taken at different simulation time snapshots of $t=17.2$, $86.0$, $172.0$, and $344.0$~$\mu^{-1}$ shown in panels (a--d) for $\theta_r=30^\circ$,  respectively. 
Flavor conversions are triggered by the initial perturbations at the center of the box [panel (a)]. 
A coherent wave-like feature develops when flavor conversions reach the nonlinear regime [panel (b)]. 
When flavor waves interact as they cross the periodic boundaries, smaller scale structures appear [panel (c)] and the system eventually settles into the final quasistationary state [panel (d)].}
\label{fig:ree_t_volume_render}
\end{figure*}

To trigger the fast flavor instabilities, we assign spherical Gaussian perturbations centered at the origin of the coordinates to $\vR_{\bm v}$ and $\bar\vR_{\bm v}$ at $t=0$ with 
\begin{subequations}
\begin{align}
\vR^0_{ee} & =\bar\vR^0_{ee}=
\frac{1}{2}
\left[ 1 + \sqrt{1-\epsilon^2(\bm x)}\right], \\
\vR^0_{xx} & =\bar\vR^0_{xx}=
\frac{1}{2}
\left[ 1 - \sqrt{1-\epsilon^2(\bm x)}\right], \\
\vR^0_{ex}&=\bar\vR^0_{ex}=\epsilon(\bm x)/2, 
\end{align}
\end{subequations}
where $\epsilon(\bm x) = 10^{-3}\exp[-(x^2 + y^2 + z^2)/2\sigma_r^2]$, and $\sigma_r=\sqrt{5}$~$\mu^{-1}$.
All simulations are performed up to $t=344$~$\mu^{-1}$ when the systems have reached the coarse-grained quasistationary states.

\section{Flavor Evolution, quasitationary state, and near flavor equilibration}\label{sec:results_numerical}

\subsection{Results in the original coordinates}\label{sec:results_unrot}

We first discuss the evolution of the system in the original coordinates. 
Fig.~\ref{fig:ree_t_volume_render}
shows the volume rendering of the phase-space averaged $\rho_{ee}$, 
defined as $\langle\rho_{ee}(\bm x)\rangle_{\Gamma}=[\int d\Gamma g_{\nu}(\bm v)\rho_{ee}(\bm v,\bm x)]$ 
at different time snapshots of 
$t =17.2$, $86.0$, $172.0$, and $344.0$~$\mu^{-1}$ for the case with $\theta_r=30^\circ$. 
Initially, the flavor instability drives the growth and spatial drift of the 3D Gaussian perturbation in the linearized regime [panel (a)]. 
When the system enters the nonlinear regime, flavor waves form and propagate, demonstrated by the coherent pattern shown in panel (b). 
At later times, when flavor waves interact, smaller-scale structures appear, as shown in panel (c). 
Eventually, when the system settles to the quasistationary state, fully developed flavor depolarization results in spatial fluctuation dominated by the length scale of $\sim 5$~$\mu^{-1}$ in the entire simulation box [panel (d)]. 
The overall behavior is qualitatively very similar to what were reported in the earlier studies (e.g.,~\cite{wu2021collective,richers2021neutrino}). 

\begin{figure}[t]
\centering   \includegraphics[width=\linewidth]{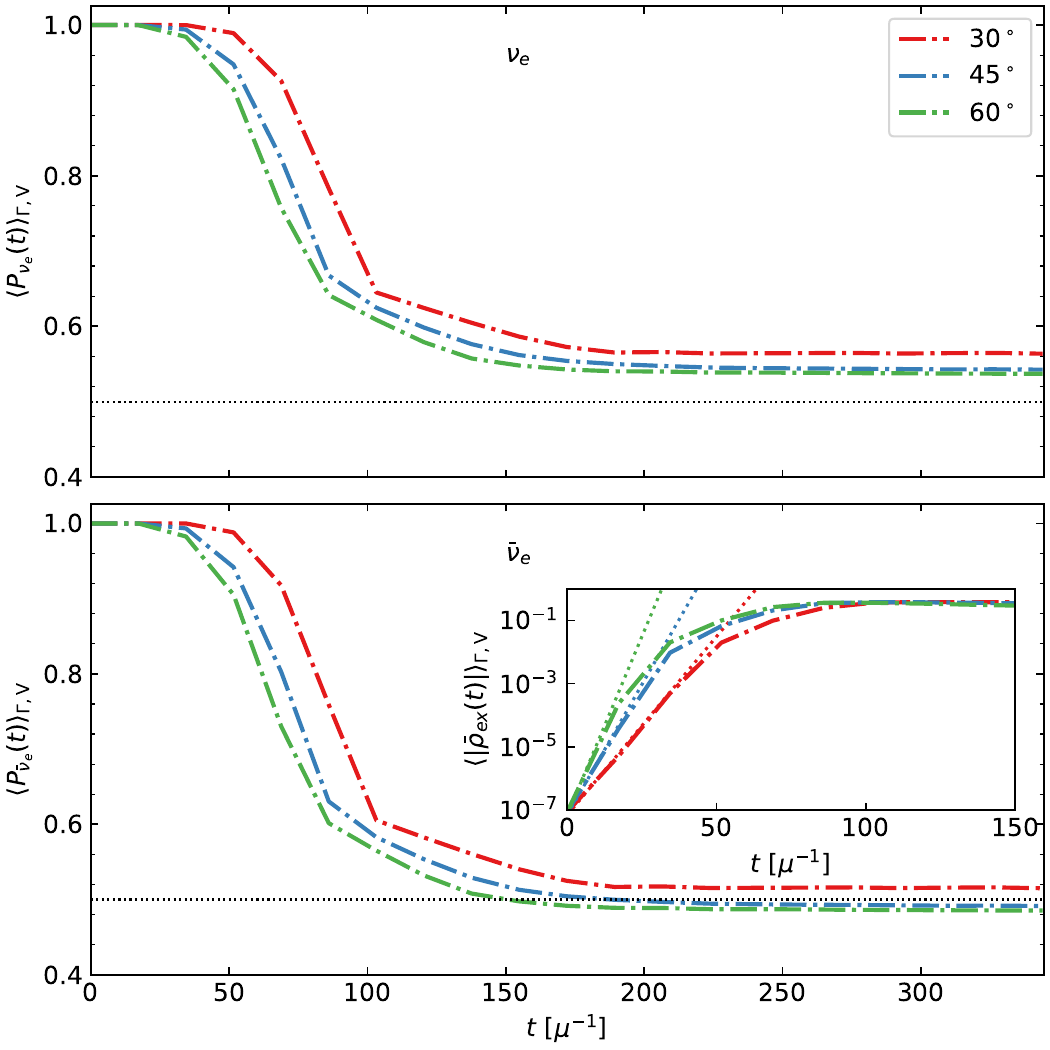}
\caption{Survival probabilities averaged over the box and the phase-space volume, $\langle P_{\nu_e} \rangle_{\Gamma,V}$ (upper panel) and 
$\langle P_{\bar\nu_e} \rangle_{\Gamma,V}$ (lower panel) as functions of time for $\theta_r=30^\circ$ (red lines), $45^\circ$ (blue lines), and $60^\circ$ (green lines).
We also show the evolution of $\langle |\bar\vR_{ex}|(t) \rangle_{\Gamma,V}$ in logarithmic scale for $t<150$~$\mu^{-1}$ in the inset of the bottom panel for different $\theta_r$ along with the maximally unstable growth rates (dotted lines) obtained from the linear stability analysis.
}
\label{fig:Pee_unrot}
\end{figure}

\begin{figure}[t]
\centering
\includegraphics[width=1.0\linewidth]{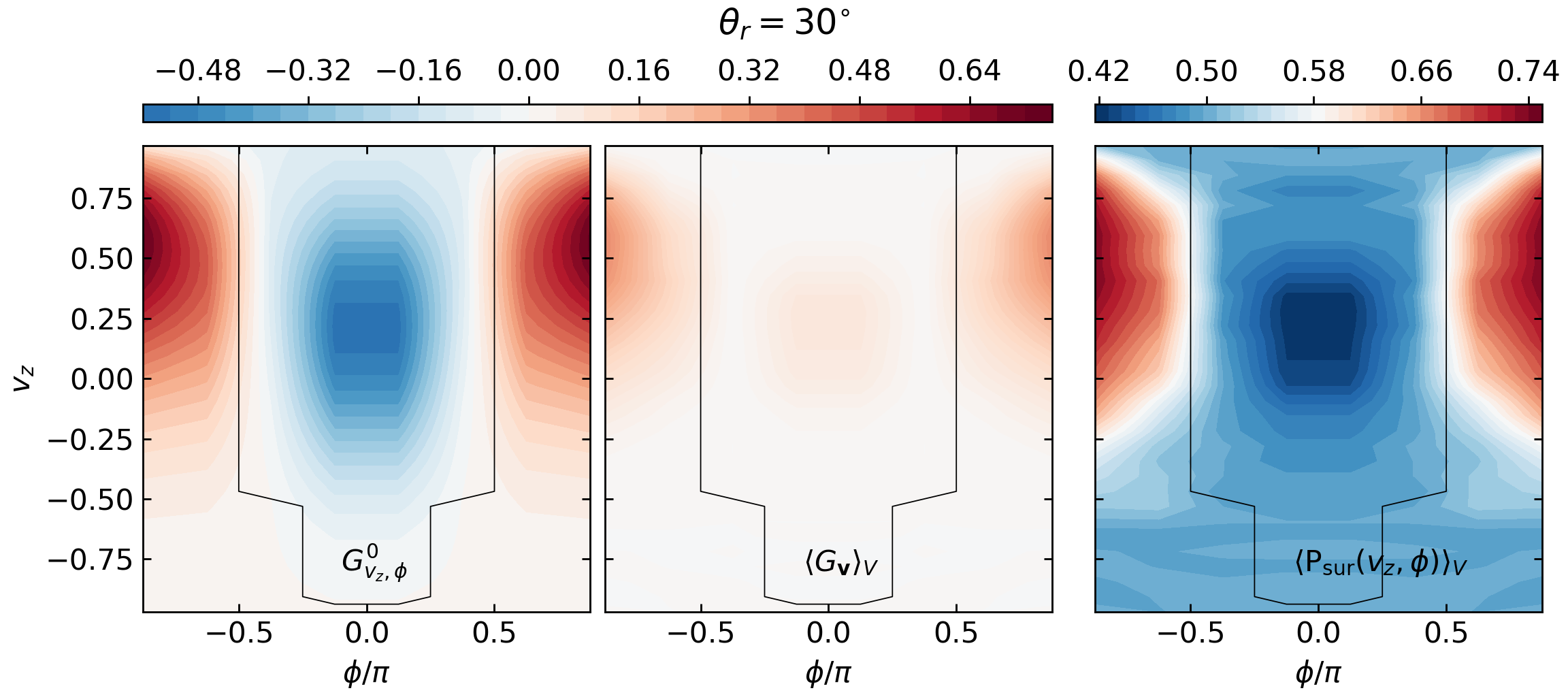}
\includegraphics[width=1.0\linewidth]{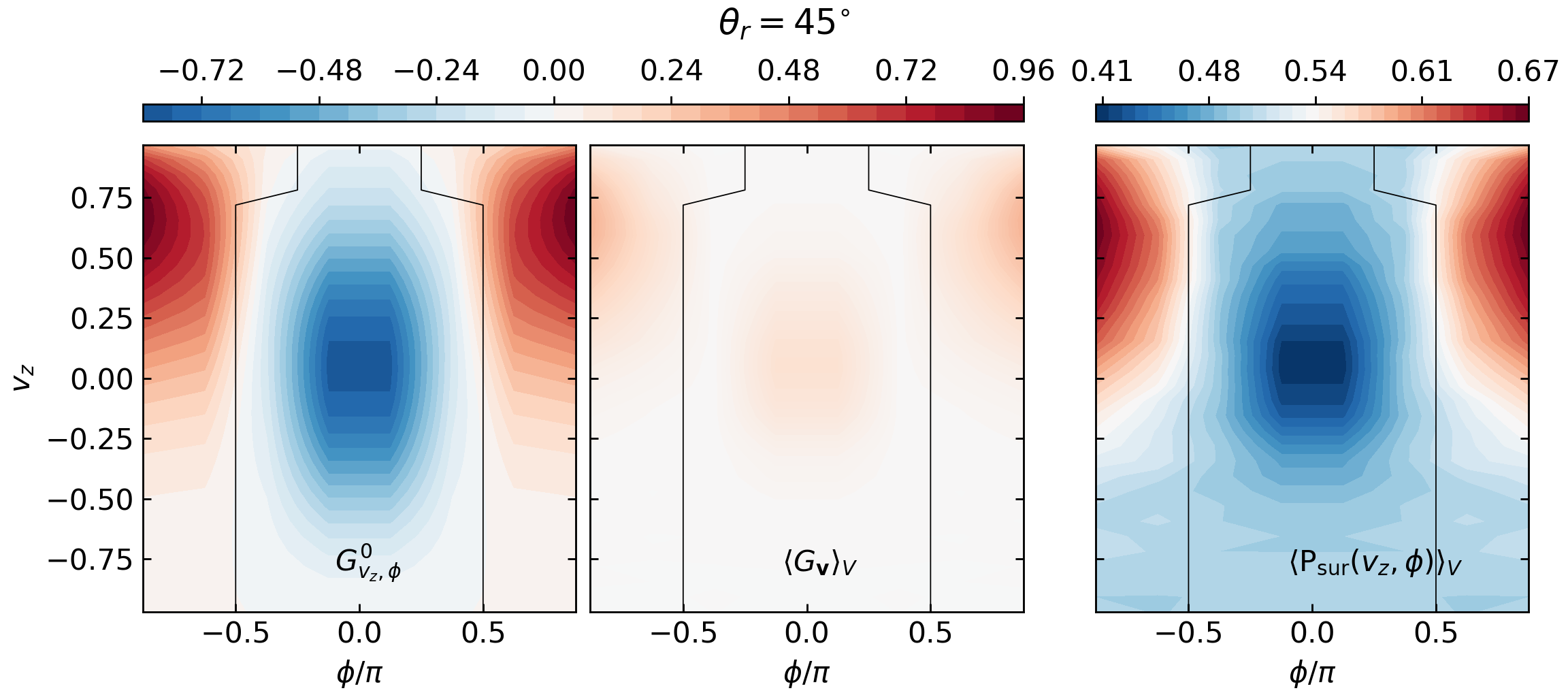}
\includegraphics[width=1.0\linewidth]{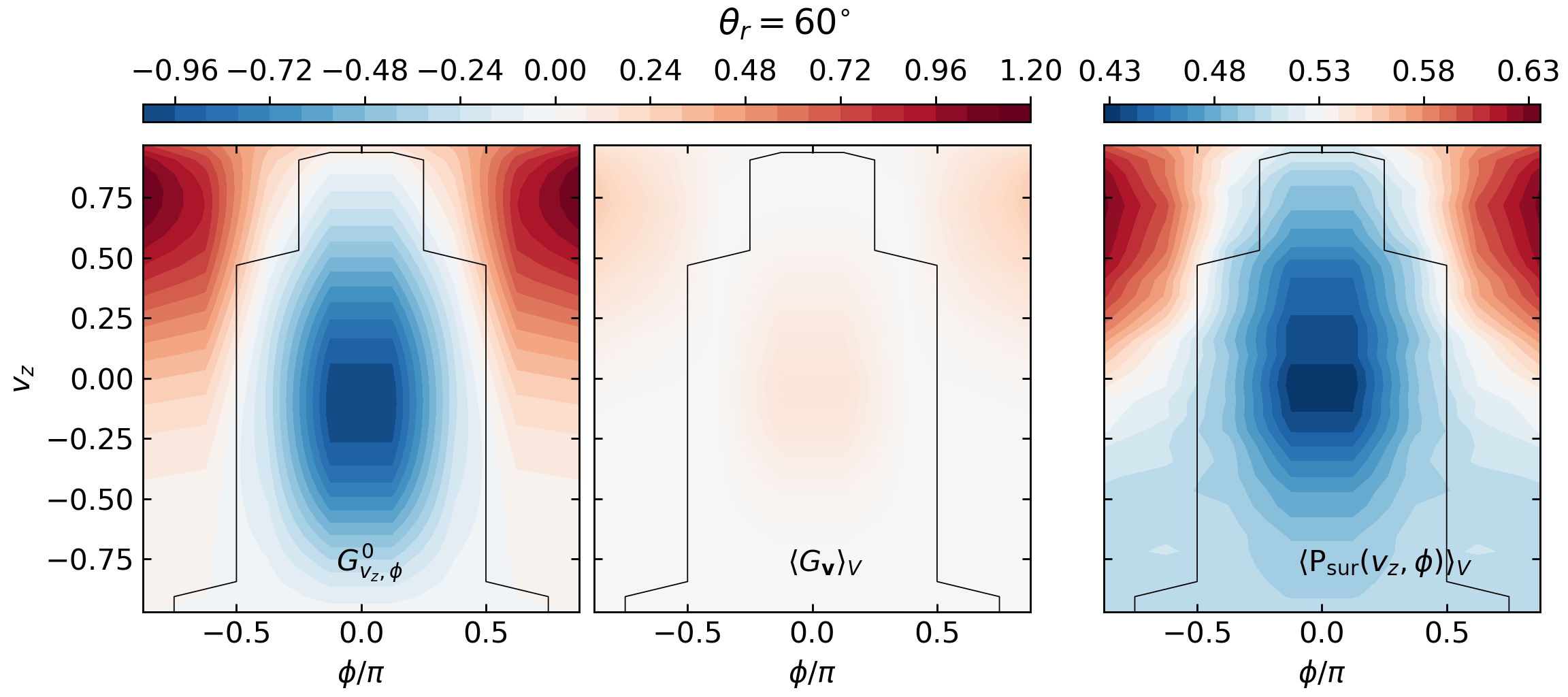}
\caption{
The initial ELN angular distribution $G^0_{v_z,\phi}$ (left panels), the spatially averaged final ELN angular distribution $\langle G_{\bm v}\rangle_V$ at $t=344~\mu^{-1}$ (middle panels), and the spatially averaged flavor survival probabilities $\langle P_{\rm sur}(v_z,\phi) \rangle_V$ (right panels) for $\theta_r=30^\circ$ (upper panels), $45^\circ$ (middle panels), and $60^\circ$ (right panels), respectively.
}
\label{fig:ELN_unrot}
\end{figure}

We show in Figure~\ref{fig:Pee_unrot} the time evolution of the $\nu_e$ and $\bar\nu_e$ survival probabilities averaged over both the 3D box volume $V$ and the phase space volume $\Gamma$ 
\begin{subequations}
\begin{align}
\langle P_{\nu_e}(t) \rangle_{\Gamma,V} & =\left [\int d\Gamma d^3{x} g_\nu(\bm v)\vR_{ee}(\bm v, \bm x, t) \right ] \bigg / L^3,\\
\langle P_{\bar\nu_e}(t) \rangle_{\Gamma,V} & =\left [\int d\Gamma d^3{x} g_{\bar\nu}(\bm v)\bar\vR_{ee}(\bm v, \bm x, t) \right ] \bigg / (\alpha L^3),
\end{align}
\end{subequations}
and in Figure~\ref{fig:ELN_unrot} the initial ELN angular distributions $G^0_{v_z,\phi}$ (left panels) as well as the final ELN distributions $\langle G_{\bm v}\rangle_V=\int d^3{x} G_{\bm v}(x)/L^3$ averaged over the entire box (middle panels) for all three cases with $\theta_r=30^\circ$, $45^\circ$, and $60^\circ$.  
Fig.~\ref{fig:Pee_unrot} shows that when taking a larger value of $\theta_r$, the increased amount of axial asymmetry leads to an earlier onset of flavor conversion as well as smaller quasistationary values of $\langle P_{\nu_e(\bar\nu_e)}(t) \rangle_{\Gamma,\bm x}$. 
This is related to the absolute values of the positive and negative parts of the initial ELN, defined by 
\begin{subequations}
\begin{align}
& I_+ = 2\pi \int d\Gamma \Theta (G^0_{v_z,\phi})G^0_{v_z,\phi},\\
& I_- = 2\pi \left |  \int d\Gamma  \Theta(-G^0_{v_z,\phi})G^0_{v_z,\phi}\right |,   
\end{align}
\end{subequations}
where $\Theta$ is the Heaviside function. 
The values of $I_+(I_-)$ are 
1.61(0.98), 2.18(1.55), and 2.76(2.13) for $\theta_r=30^\circ$, $45^\circ$, and $60^\circ$, respectively. 
Note that $I_+ - I_- = 2\pi(1-\alpha)$ is the same for all cases. 
Clearly, the more asymmetric cases (with larger $\theta_r$) have larger values of $I_+$ and $I_-$, which can also be seen from the initial ELN distributions shown in the left panels of Fig.~\ref{fig:ELN_unrot}. 
When $I_+$ and $I_-$ are larger, the associated instability growth rate is also larger (see the inset of Fig.~\ref{fig:Pee_unrot} showing the evolution of $\langle |\bar\vR_{ex}|(t)\rangle_{\Gamma,V}$ in logarithmic scale for $t<150$~$\mu^{-1}$)\footnote{The dotted lines in the inset of Fig.~\ref{fig:Pee_unrot} are the respective maximally unstable growth rates obtained from the linear stability analysis~\cite{banerjee2011linearized,raffelt2013axial} taking the same angular grids for $|k_{x,y,z}/\mu|<\pi$.}, resulting in the earlier onset of flavor conversions. 
What is also similar to the 1D box cases is that the ELN crossing is erased when averaging over the entire 3D box, as shown in the middle panels of Fig.~\ref{fig:ELN_unrot}. 
Since the total ELN in the box is conserved with the periodic boundary condition, more flavor conversions happen when $I_-$ is larger, resulting in a lower value of the quasistationary $\langle P_{\nu_e(\bar\nu_e)} \rangle_{\Gamma,V}$. 

Besides the elimination of the ELN crossing, the middle panels of Fig.~\ref{fig:ELN_unrot} also show that for the angular region where $G_{v_z,\phi}^0<0$ initially, the corresponding values of  
$\langle G_{\bm v}\rangle_V=\int d^3{x} G_{\bm v}(x)/L^3$ become nearly zero, showing that near flavor equilibration on the coarse-grained level in most part of this angular domain is reached. 
For completeness, we show in the right panels of Fig.~\ref{fig:ELN_unrot} the final coarse-grained angle-dependent flavor survival probabilities 
$\langle P_{\rm sur}(v_z,\phi)\rangle_V = \int d^3{x} \rho_{ee}(\bm x, \bm v)/L^3$ taken at the final time snapshot for all three cases. 
Once again, these panels confirm that $\langle P_{\rm sur}(v_z,\phi)\rangle_V\simeq 0.5$ is obtained in the angular domain whose $G_{v_z,\phi}^0<0$. 
They also show more clearly that slight flavor overconversion with $\langle P_{\rm sur}(v_z,\phi)\rangle_V\lesssim 0.5$ happens in regions where $G_{v_z,\phi}^0$ are more negative around $\phi\sim 0$. 
All these suggest that similar conclusions derived based on the 1D box simulations hold in a more general setting where both the axisymmetry in the angular distribution and the translation symmetry in spatial dimensions are broken explicitly. 
We note that for all cases the initial reflection symmetry with respect to $\phi \rightarrow -\phi$ is preserved throughout the evolution for all cases as can be inferred from Fig.~\ref{fig:ELN_unrot}. 

\begin{figure}[t]
\centering
\includegraphics[width=1.0\linewidth]{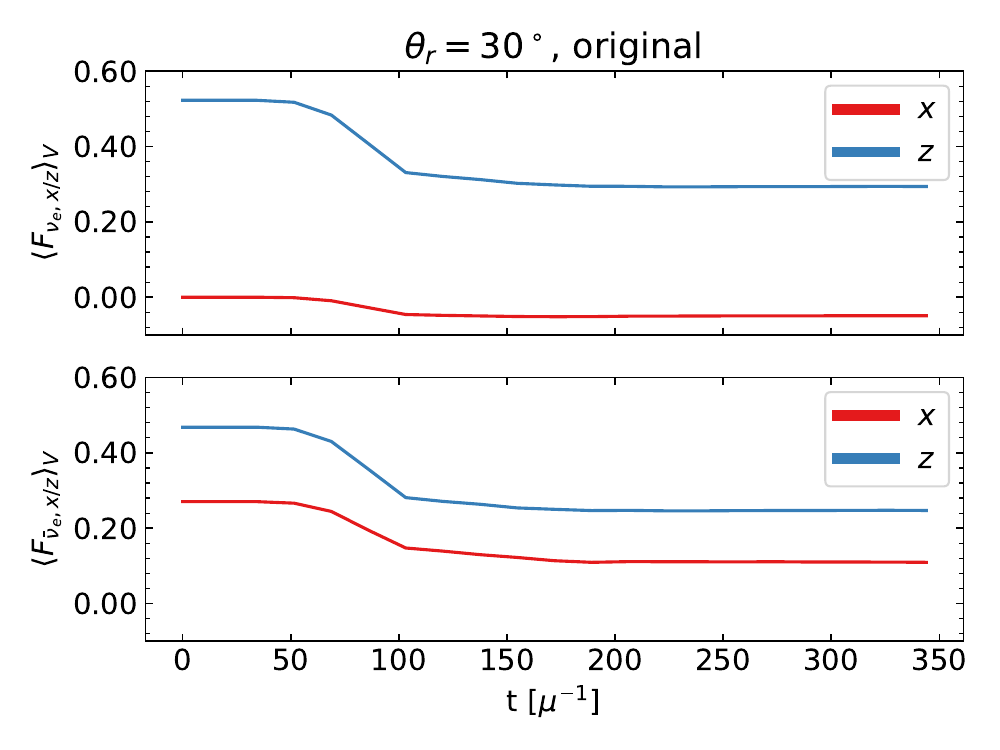}
\caption{Time evolution of the $x$ (red curves) and $z$ (blue curves) components of $\langle\bm F_{\nu_e}\rangle_V$ (upper panel) and $\langle\bm F_{\bar\nu_e}\rangle_V$ (lower panel) for $\theta=30^\circ$. 
}
\label{fig:moments_unrot}
\end{figure}

Figure~\ref{fig:moments_unrot} shows the evolution of the spatially averaged $x$ and $z$ components of the $\nu_e$ and $\bar\nu_e$ first angular moments (normalized by $n_\nu$)  
\begin{subequations}
\begin{align}
\langle \bm F_{\nu_e} \rangle_V
& =\frac{1}{L^3}\int d^3{x} d\Gamma  \bm v g_\nu\vR_{\bm v}(x), \\
\langle \bm F_{\bar\nu_e} \rangle_V
& = \frac{1}{L^3} \int d^3{x} d\Gamma  \bm v g_{\bar\nu}\bar\vR_{\bm v}(x),
\end{align}
\end{subequations}
for $\theta_r=30^\circ$. 
For the initially nonzero flux vector components ($\langle \bm F_{\nu_e,z} \rangle_V$, $\langle \bm F_{\bar\nu_e,z} \rangle_V$, and $\langle \bm F_{\bar\nu_e,x} \rangle_V$), 
flavor conversions of $\nu_e$ ($\bar\nu_e$) to $\nu_x$ ($\bar\nu_x$) lead to the reduction of their 
values in all cases.  
For $\langle F^0_{\nu_e,x}\rangle_V$, which is initially zero,  
since FFC converts more (less) $\nu_e$ to $\nu_x$ around $\phi\sim 0$ ($\phi\sim \pi$) as shown in the right panels of Fig.~\ref{fig:ELN_unrot}, it 
becomes negative over time  
as a result of the broken axisymmetry. 
Similar evolution of the flux vector components are obtained for other values of $\theta_r$. 

\subsection{Results in the Rotated Coordinates}\label{sec:result_rotated}
We now turn our attention to results obtained in the rotated coordinates. 
Since the physical evolution of the system is independent of the choice of the coordinate system, we do not repeat the flavor evolution and the properties of the quasistationary states reported in Sec.~\ref{sec:results_unrot}. 
Instead, we focus on comparing results obtained by taking the axially asymmetric ELN distribution $G^0_{v'_z,\phi'}$ and the corresponding axisymmetric ELN distribution $G^{a,0}_{v'_z}=\int d\phi' G^0_{v'_z,\phi'} / (2\pi)$, constructed by averaging over $\phi'$ from $G^0_{v'_z,\phi'}$. 
Since this coordinate system is chosen such that the only nonzero initial first ELN angular moment 
$\int d\Gamma' \bm v G^0_{v'_z,\phi'}$ is along the $z'$ direction as introduced in Sec.~\ref{subsec:setup::angdistr}, the initial zeroth and first ELN angular moments evaluated using axisymmetric ELN  $G^{a,0}_{v'_z}$ are identical to those evaluated using $G^0_{v'_z,\phi'}$. 

\begin{figure}[t]
\centering
\includegraphics[width=1.0\linewidth]{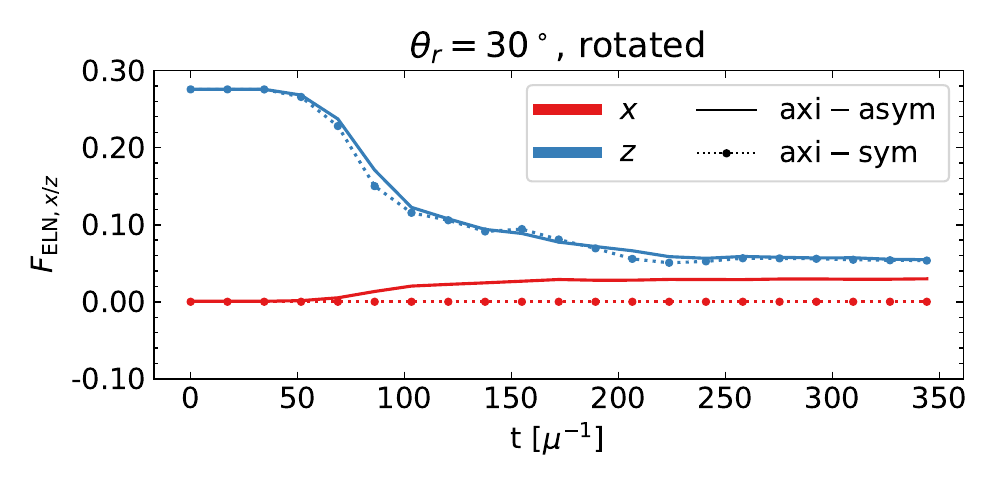}
\caption{
Evolution of the first ELN angular moments obtained in the rotated coordinates with axially asymmetric initial distribution function ($F_{{\rm ELN},x/z}$, solid curves) and with axisymmetric distribution ($F^a_{{\rm ELN},x/z}$, dotted curves with dots) for $\theta_r=30^\circ$. 
}
\label{fig:elnmoments_rot}
\end{figure}

Figure~\ref{fig:elnmoments_rot} compares the time evolution of the 
first ELN moments (averaged over the box volume) for the axially asymmetric (${\bm F}_{\rm ELN}$) and the axisymmetric (${\bm F}^a_{\rm ELN}$) cases for $\theta_r=30^\circ$, defined as 
\begin{subequations}
\begin{align}
{\bm F}_{\rm ELN} & = \frac{1}{L^3} \int d^3{x'} 
 d\Gamma {\bm v'} G_{v'_z,\phi'}, \\
 {\bm F}^a_{\rm ELN} & = \frac{1}{L^3} \int d^3{x'} 
 d\Gamma {\bm v'} G^a_{v'_z}. 
\end{align}
\end{subequations}
It shows that the evolution of 
$F_{{\rm ELN}, z'}$ and 
$F^a_{{\rm ELN}, z'}$ closely follow each other. 
However, for the axially asymmetric cases, 
$F_{{\rm ELN},x'}$ becomes nonzero due to the breaking of the axisymmetry, while $F^a_{{\rm ELN}, x'}$ remains zero throughout the evolution. 

\begin{figure}[t]
\centering
\includegraphics[width=1.0\linewidth]{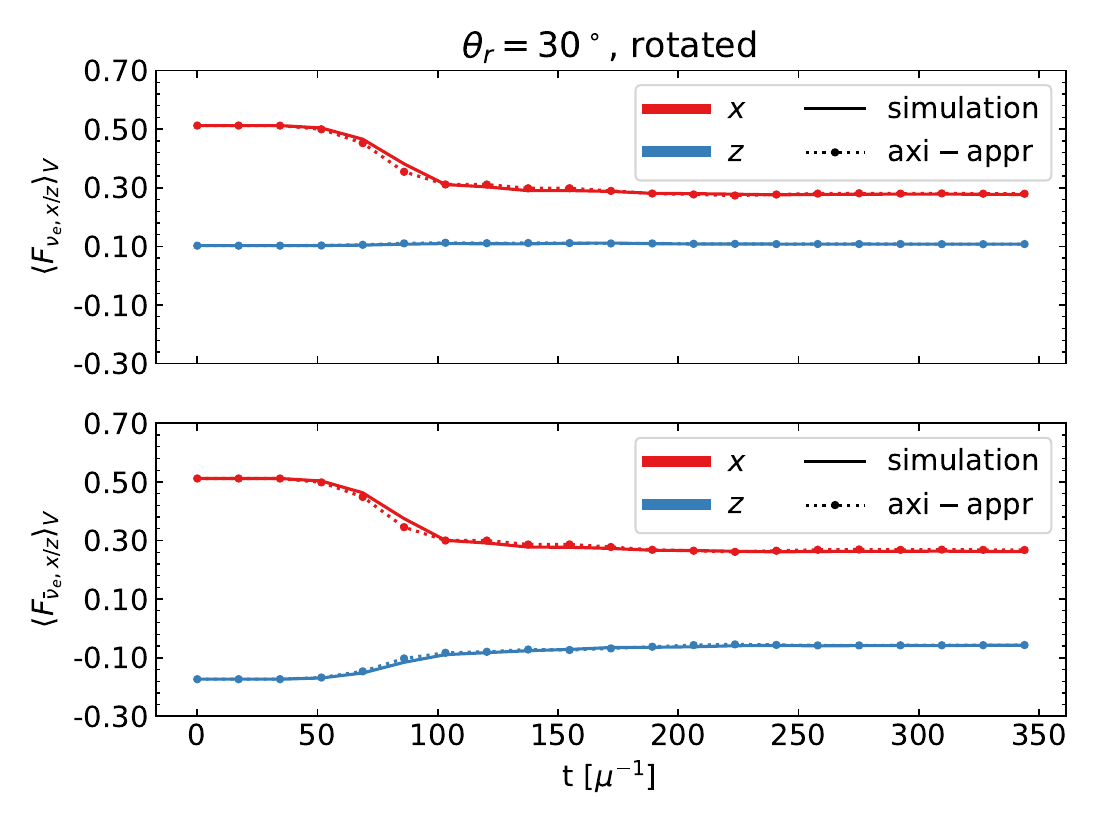}
\caption{
Comparison of the evolution of the $\nu_e$ (upper panel) and $\bar\nu_e$ (lower panel) flux vector components obtained from simulations with axially asymmetric distribution (solid curves) with that estimated using coarse-grained survival probabilities obtained in the corresponding axisymmetric simulations (dotted curves with dots) [see Eq.~\eqref{eq:axi-approx-moments}] in the rotated coordinates with $\theta_r=30^\circ$. 
}
\label{fig:momentz_rot}
\end{figure}

Interestingly, if we take the time-dependent, spatially averaged flavor survival probability computed in the axially symmetric cases $\langle P^{\rm axi}_{ee}(t,v_z')\rangle$ and use it together with the angular distribution functions used in the axially asymmetric cases to approximately evaluate the time evolution of the survival probability, flux-ratio vectors, or ELN moments, it results in remarkable agreements. 
Figure~\ref{fig:momentz_rot} shows the comparison of 
$\langle F_{\nu_e,x'}\rangle_{V'}$, $\langle F_{\nu_e,z'}\rangle_{V'}$, $\langle F_{\bar\nu_e,x'}\rangle_{V'}$, $\langle F_{\bar\nu_e,z'}\rangle_{V'}$ from the axially asymmetric simulations with those evaluated by 
\begin{equation}\label{eq:axi-approx-moments}
{\bm F}_{\nu_e(\bar\nu_e)}^{\rm axi-appr}  = \int d\Gamma' \bm v' g_{\nu(\bar\nu)}(\bm v') \langle P_{ee}^{\rm axi}(t,v_z') \rangle_{V'}, 
\end{equation}
for the case with $\theta_r=30^\circ$ as an example. 
It clearly shows that one can use $\langle P^{\rm axi}_{ee}(t,v_z')\rangle_{V'}$ derived in the axially symmetric simulations in the rotated frame with the axially asymmetric initial angular distributions to closely predict the time evolution of the angle-integrated quantities. 
We have also verified that similar agreements apply to cases with different values of $\theta_r$. 

\section{Approximated prescriptions for evaluating the angular moments}\label{sec:approx_prescript}

A main goal of conducting local box FFC simulations is to find simple prescriptions to characterizing the post-FFC angular distributions or angular moments for practical implementation into effective classical neutrino transport models~\cite{xiong2024robust}.  
Below, we discuss how to generalize different analytical prescriptions proposed in \cite{zaizen2023simple,xiong2023evaluating} for 3D cases without axisymmetry in neutrino angular distributions.  

First, given the fact that near flavor equilibrium is reached in one of the angular domain, it is straightforward to generalize the prescription of the ``boxlike'' scheme~\cite{zaizen2023simple,xiong2023evaluating} to approximate the post-FFC survival probabilities for both $\nu_e$ and $\bar\nu_e$ with 
\begin{equation}\label{eq:box_presc}
P^b_{\rm sur}(v_z,\phi) =
\begin{cases}
\frac{1}{2} & {\rm for~}\Gamma^<, \\
1-I_</(2 I_>) & {\rm for~}\Gamma^>,
\end{cases}
\end{equation}
where $I_<=\min(I_-,I_+)$,  $I_>=\max(I_-,I_+)$ and $\Gamma^<$ ($\Gamma^>$) denotes the corresponding angular domain that contributes to $I_<$ ($I_>$). 
Equation~\eqref{eq:box_presc} allows us to approximately evaluate various coarse-grained angle-integrated quantities after FFC has settled to the quasistationary state in a straightforward manner. 
For instance, the 
spatially averaged post-FFC zeroth and first angular moments in this boxlike scheme can be estimated by 
\begin{subequations}\label{eq:moments_appro}
\begin{align}
N^p_{\nu_e} & = \int d\Gamma g_{\nu}(\bm v) P^b_{\rm sur}(v_z,\phi),  \\
N^p_{\bar\nu_e} & = \int d\Gamma g_{\bar\nu}(\bm v) P^b_{\rm sur}(v_z,\phi), \\
\bm F^p_{\nu_e} & = \int d\Gamma \bm v g_{\nu}(\bm v) P^b_{\rm sur}(v_z,\phi),  \\
\bm F^p_{\bar\nu_e} & = \int d\Gamma \bm v g_{\bar\nu}(\bm v) P^b_{\rm sur}(v_z,\phi). 
\end{align}
\end{subequations}

\begin{table}
\begin{tabular}{||c|c||c|c|c|c|c|c||}
\hline
$\theta_r$ &
Scheme & 
$N^p_{\nu_e}$ &
$N^p_{\bar\nu_e}$ & 
$F^p_{\nu_e,x}$ & 
$F^p_{\nu_e,z}$ & 
$F^p_{\bar\nu_e,x}$ & 
$F^p_{\bar\nu_e,z}$
\\ 
\hline
$30^\circ$ & sim         & 0.563 & 0.463 & -0.049 & 0.294 & 0.109 & 0.247 \\
$30^\circ$ & box         & 0.599 & 0.498 & -0.047 & 0.312 & 0.118 & 0.267 \\
$30^\circ$ & power-1/2-s & 0.583 & 0.483 & -0.050 & 0.302 & 0.120 & 0.256 \\
$30^\circ$ & power-1/2-a & 0.570 & 0.470 & -0.047 & 0.285 & 0.125 & 0.244 \\
\hline 
$45^\circ$ & sim         & 0.542 & 0.442 & -0.036 & 0.288 & 0.168 & 0.197 \\
$45^\circ$ & box         & 0.584 & 0.484 & -0.033 & 0.309 & 0.186 & 0.217 \\
$45^\circ$ & power-1/2-s & 0.566 & 0.466 & -0.039 & 0.297 & 0.185 & 0.203 \\
$45^\circ$ & power-1/2-a & 0.559 & 0.459 & -0.041 & 0.283 & 0.187 & 0.196 \\
\hline 
$60^\circ$ & sim         & 0.537 & 0.436 & -0.027 & 0.288 & 0.213 & 0.141 \\
$60^\circ$ & box         & 0.576 & 0.476 & -0.023 & 0.306 & 0.235 & 0.155 \\
$60^\circ$ & power-1/2-s & 0.559 & 0.459 & -0.032 & 0.296 & 0.232 & 0.143 \\
$60^\circ$ & power-1/2-a & 0.555 & 0.455 & -0.036 & 0.286 & 0.232 & 0.138 \\
\hline
\end{tabular}
\caption{
Values of the post-FFC spatially averaged $\nu_e$ and $\bar\nu_e$ zeroth angular moments as well as their $x$ and $z$ components of the first angular moments obtained from numerical simulation (scheme ``sim''), estimated using the box-like prescription for survival probability [scheme ``box''; see Eq.~\eqref{eq:box_presc}], the power-1/2-s prescription [scheme ``power-1/2-s''; see Eq.~\eqref{eq:p12_presc}], and the power-1/2-a prescription [scheme ``power-1/2-a''; see Eq.~\eqref{eq:p12a_presc}]. 
} 
\label{table:final_moments}
\end{table}	

Second, the excellent agreement in the evolution of moments shown in Fig.~\ref{fig:momentz_rot} in the rotated coordinates 
suggests that one may directly utilize the improved analytical prescription developed based on axisymmetric simulations \cite{xiong2023evaluating} to approximately evaluate the post-FFC angular moments. 
Below, we show how a axisymmetric (in the rotated coordinates) prescription can be directly  
employed in the multidimensional condition. 
First, we find the angular domains corresponding to $I_<$ and $I_>$ as small and large sides, respectively.
The flavor equilibration is assumed on the small side.
For the large side, we will determine the distribution of survival probability as follows.
We find the initial ELN flux vector 
$\bm F^0_\mathrm{ELN}$ whose  
direction points to where the large side is located.
We then compute the values of the projection of all ELN crossing velocity $\bm v_c$ along ${\bm{\hat{F}}}^0_\mathrm{ELN}$ and find the maximal projected value as $v_c^m$, where $\bm F^0_\mathrm{ELN}=\bm F^0_\mathrm{ELN}/|\bm F^0_\mathrm{ELN}|$. 
The survival probability on the large side is then given by
\begin{equation}\label{eq:p12_presc}
P^{p-s}_{\rm sur}(\bm v)=
\begin{cases}
\frac{1}{2}, &  {\rm for~}\bm v \cdot {\bm{\hat{F}}}^0_\mathrm{ELN} \leq v_c^m, \\
1-\frac{1}{2} h(\frac{|\bm v \cdot {\bm{\hat{F}}}^0_\mathrm{ELN} -v_c^m|}{a} ), &  {\rm for~}\bm v \cdot {\bm{\hat{F}}}^0_\mathrm{ELN} >v_c^m,
\end{cases}
\end{equation}
where $h(x)=(x^2+1)^{-1/2}$, and the parameter $a$ can be determined using the following equation derived from the conservation of the ELN,  
\begin{equation}\label{eq:ELNconser_a}
I_< = \int_{\Gamma^>} d\Gamma G(\bm v) h(|\bm v \cdot {\bm{\hat{F}}}^0_\mathrm{ELN} -v_c^m|/a ),
\end{equation}
where the integration is performed over the domain $\Gamma^>$, 
generalized from \cite{xiong2023evaluating}\footnote{The integrand in the right-hand side of Eq.~\eqref{eq:ELNconser_a} is a monotonic function of $a$, so a unique solution can always be found.}.
We then use Eq.~\eqref{eq:p12_presc} to replace $P^b_{\rm sur}(\bm v)$ in Eqs.~\eqref{eq:moments_appro}
to compute the corresponding values of the zeroth and the first moments. 
This scheme is named ``power-1/2-s''.  

Third, we can further generalize the power-1/2 prescription to include the axial asymmetric feature of the results, and use it to evaluate the post-FFC moments, dubbed as the ``power-1/2-a'' scheme.
Just as the original motivation for the power-1/2 prescription in 1D-box simulations, which is to provide a prescription with the survival probability smoothly transitioning across the angular crossing, we achieve the same goal for the power-1/2-a scheme,  by defining a shortest  ``distance'' to the initial ELN crossing contour as  
\begin{equation}
    D(\bm v) = 1-f_{\rm max}(\bm v),
\end{equation}
where the function $f_{\rm max}(\bm v)$ returns the maximal value of $\bm v \cdot \bm v_c$ for all $\bm v_c$ with a given $\bm v$. 
The corresponding survival probability to predict the quasistationary state in this scheme is then defined by
\begin{equation}\label{eq:p12a_presc}
P^{p-a}_{\rm sur}(\bm v)=
\begin{cases}
\frac{1}{2}, &  {\rm for~}\Gamma^<, \\
1-\frac{1}{2} h[D(\bm v)/a], &  {\rm for~}\Gamma^>. 
\end{cases}
\end{equation}
The determination of $a$ in the power-1/2-a scheme follows the 
same procedure outline above for the 
the power-1/2-s scheme.

\begin{figure*}[t]
\centering
\includegraphics[width=.3\textwidth]{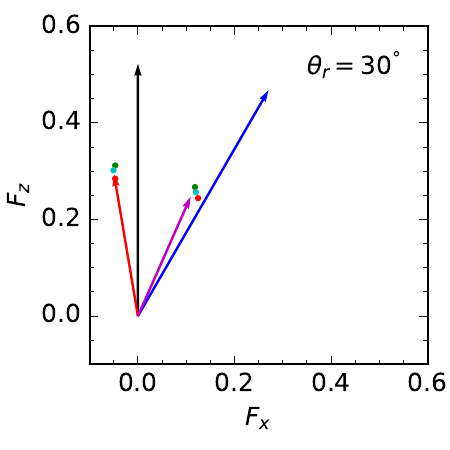}
\includegraphics[width=.3\textwidth]{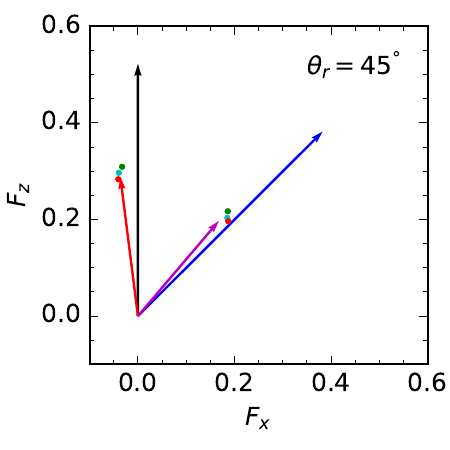}
\includegraphics[width=.3\textwidth]{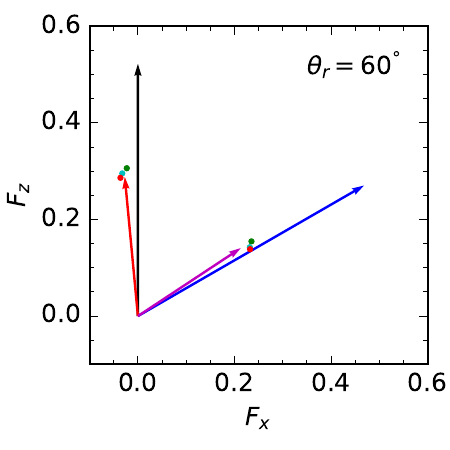}
\caption{
Graphical representation of the 
initial $\nu_e$ (black arrows) and $\bar\nu_e$ flux vectors (blue arrows), their final coarse-grained vectors from simulations (red and magenta arrows), and the values predicted using the box-like (green dots), the power-1/2-s prescription (cyan dots), as well as the power-1/2-a prescription (red dots) for $\theta_r=30^\circ$ (left panel), $45^\circ$ (middle panel), and $60^\circ$ (right panel) in the original coordinates, respectively.
\label{fig:fluxvec_unrot}}
\end{figure*}

Table~\ref{table:final_moments}
compares the values of the $\nu_e$ and $\bar\nu_e$ zeroth moments as well as the $x$ and $z$ components of their first moments derived from the simulations in the original coordinates to those evaluated using the above three different prescriptions for all three different $\theta_r$ values. 
In all cases, the fractional differences between values obtained by numerical simulations and those with prescriptions are smaller than $\sim 10 \%$. 
Unsurprisingly, the boxlike scheme generally leads to larger errors, due to its simplest form that is discontinuous. 
For the power-1/2-s scheme and the power-1/2-a scheme, the fractional differences are generally smaller compared to the boxlike scheme, especially for the zeroth moment as well as the $z$ component of the flux vectors.  
Figure~\ref{fig:fluxvec_unrot} shows the graphical representation of the initial $\nu_e$ (black arrows) and $\bar\nu_e$ flux vectors (blue arrows), their final coarse-grained vectors from simulations (red and magenta arrows), and the values predicted using the box-like (green dots), the power-1/2-s prescription (cyan dots), as well as the power-1/2-a prescription (red dots) for all three cases with different $\theta_r$ to demonstrate the impact of FFC, and the difference between the numerical outcome and the approximated evaluations. 
Once again, it illustrates that in general the power-1/2-a scheme provides the best prediction for the post-FFC flux vectors, followed by the power-1/2-s scheme, and then the boxlike scheme.  

Finally, we note that in the power-1/2-s scheme that assumes axisymmetry in the rotated coordinates we have conservatively chosen the maximally projected velocity $v_c^m$ as the critical velocity value in Eq.~\eqref{eq:p12_presc}. 
This is to avoid having a final ELN angular distribution that still contains ELN crossings. 
If we have used the averaged projected velocity value as the critical velocity in Eq.~\eqref{eq:p12_presc}, it will result in similar approximated numbers for the final moments, but the corresponding final ELN angular distribution will have crossings. 
Such a choice can lead to inconsistency when implemented into the effective classical transport model generalized from \cite{xiong2024robust} in multidimensions.
We also caution that this method is not guaranteed to work for very extreme cases where $\bm v_c\cdot {\bm{\hat{F}}}^0_\mathrm{ELN}$ span a very wide range of values. 
As for the further improved power-1/2-a scheme, it does not suffer from this issue and can be universally used for any ELN distributions without axisymmetry. 
Also noted is that for cases with very shallow ELN angular crossings, the power-1/2-a scheme automatically converges toward the boxlike scheme.

\section{Conclusion and outlook}
\label{sec:concluion}
In this work, we have conducted numerical simulations of collective neutrino fast flavor conversions in three spatial dimensions in a cubic box with periodic boundary conditions using the extended version of \textsc{cose}$\nu$. 
We first examined three different physical cases with axially asymmetric neutrino ELN angular distributions in a coordinate system where the $x$ and $y$ components of the $\nu_e$ flux vector are zero. 
Our results show that when the system has reached the quasistationary state, the spatially averaged ELN angular crossing is erased, and near flavor equilibration is reached in one angular domain defined by the initial ELN crossing contour, similar to what previously observed in the 1D box cases. 
We have further conducted the same set of axially asymmetric simulations in rotated coordinate systems where the $x'$ and $y'$ components of the ELN angular moments are zero, as well as the corresponding axisymmetric counterparts. 
We found that the evolution of the zeroth ELN moments and the $z'$ components of the first ELN moments in these two sets of simulations are nearly identical, while the evolution of the $x'$ components of the first ELN  moments differs.  
Interestingly, we find that by folding the spatially averaged flavor survival probabilities computed based the axisymmetric counterpart distributions with the initially axially asymmetric angular distributions. 

Based on the simulation outcomes, we have provided a generalized boxlike scheme 
and two different versions (power-1/2-s and power-1/2-a) based on the power-1/2 scheme proposed in \cite{xiong2023evaluating} to approximate the post-FFC flavor survival probabilities in 3D. 
The generalized power-1/2-s scheme assumes axisymmetry in the rotated coordinates while the power-1/2-a scheme takes into account the axial asymmetric feature of the system.     
While we have found that all these schemes give rise to less than $\lesssim 10\%$ errors in the predicted post-FFC moments, the power-1/2-a scheme predicts the post-FFC moments more accurately than the other two approximated prescriptions.
We note that, however, the implementation of the power-1/2-a and the power-1/2-s schemes require more computation time than the simplest boxlike scheme, for their better accuracy. 

The numerical studies presented in this work indicate that the important feature -- the erasure of the spatially averaged ELN crossing and the coarse-grained flavor equilibration in an angular domain -- concluded based on 1D periodic box simulations with axisymmetric neutrino distributions remains generally valid in multidimensional cases without axisymmetry. 
Hence, it is expected that one can similarly implement the analytical prescriptions proposed in this paper into the effective classical transport model proposed in \cite{xiong2024robust} to take into account the effect of FFC in global neutrino transport simulation in multidimensions. 
This aspect will be further pursued in the future. 

\begin{acknowledgments}
M.-R.W. and M.G. acknowledge support from the National Science and Technology Council, Taiwan under Grant No.~111-2628-M-001-003-MY4, and the Academia Sinica (Project No.~AS-CDA-109-M11).
M.-R.W also acknowledges support from the Physics Division of the National Center for Theoretical Sciences, Taiwan.
Z.X. acknowledges support of the European Research Council (ERC) under the European Union’s Horizon 2020 research and innovation programme (ERC Advanced Grant KILONOVA No. 885281), the Deutsche Forschungsgemeinschaft (DFG, German Research Foundation) -- Project-ID 279384907 -- SFB 1245, and MA 4248/3-1. 
C.-Y.L. acknowledges support from the National Center for High-performance Computing (NCHC).
We acknowledge the computing resources provided by the Academia Sinica Grid-computing Center and the National Center for High Performance Computing in Taiwan. 
We acknowledge the following software: \textsc{matplotlib}~\cite{matplotlib}, \textsc{numpy}~\cite{numpy}, and \texttt{yt}~\cite{turk2010yt}. 
\end{acknowledgments}

\section*{Data Availability}
The data that support the findings of this article are openly available at \url{https://doi.org/10.5281/zenodo.14220225}.

\bibliographystyle{apsrev4-1}

\begin{thebibliography}{75}%
\makeatletter
\providecommand \@ifxundefined [1]{%
 \@ifx{#1\undefined}
}%
\providecommand \@ifnum [1]{%
 \ifnum #1\expandafter \@firstoftwo
 \else \expandafter \@secondoftwo
 \fi
}%
\providecommand \@ifx [1]{%
 \ifx #1\expandafter \@firstoftwo
 \else \expandafter \@secondoftwo
 \fi
}%
\providecommand \natexlab [1]{#1}%
\providecommand \enquote  [1]{``#1''}%
\providecommand \bibnamefont  [1]{#1}%
\providecommand \bibfnamefont [1]{#1}%
\providecommand \citenamefont [1]{#1}%
\providecommand \href@noop [0]{\@secondoftwo}%
\providecommand \href [0]{\begingroup \@sanitize@url \@href}%
\providecommand \@href[1]{\@@startlink{#1}\@@href}%
\providecommand \@@href[1]{\endgroup#1\@@endlink}%
\providecommand \@sanitize@url [0]{\catcode `\\12\catcode `\$12\catcode
  `\&12\catcode `\#12\catcode `\^12\catcode `\_12\catcode `\%12\relax}%
\providecommand \@@startlink[1]{}%
\providecommand \@@endlink[0]{}%
\providecommand \url  [0]{\begingroup\@sanitize@url \@url }%
\providecommand \@url [1]{\endgroup\@href {#1}{\urlprefix }}%
\providecommand \urlprefix  [0]{URL }%
\providecommand \Eprint [0]{\href }%
\providecommand \doibase [0]{http://dx.doi.org/}%
\providecommand \selectlanguage [0]{\@gobble}%
\providecommand \bibinfo  [0]{\@secondoftwo}%
\providecommand \bibfield  [0]{\@secondoftwo}%
\providecommand \translation [1]{[#1]}%
\providecommand \BibitemOpen [0]{}%
\providecommand \bibitemStop [0]{}%
\providecommand \bibitemNoStop [0]{.\EOS\space}%
\providecommand \EOS [0]{\spacefactor3000\relax}%
\providecommand \BibitemShut  [1]{\csname bibitem#1\endcsname}%
\let\auto@bib@innerbib\@empty
\bibitem [{\citenamefont {Pantaleone}(1992)}]{pantaleone1992neutrino}%
  \BibitemOpen
  \bibfield  {author} {\bibinfo {author} {\bibfnamefont {J.}~\bibnamefont
  {Pantaleone}},\ }\href {\doibase 10.1016/0370-2693(92)91887-F} {\bibfield
  {journal} {\bibinfo  {journal} {Phys. Lett. B}\ }\textbf {\bibinfo {volume}
  {287}},\ \bibinfo {pages} {128} (\bibinfo {year} {1992})}\BibitemShut
  {NoStop}%
\bibitem [{\citenamefont {Sigl}\ and\ \citenamefont
  {Raffelt}(1993)}]{sigl1993general}%
  \BibitemOpen
  \bibfield  {author} {\bibinfo {author} {\bibfnamefont {G.}~\bibnamefont
  {Sigl}}\ and\ \bibinfo {author} {\bibfnamefont {G.}~\bibnamefont {Raffelt}},\
  }\href {\doibase 10.1016/0550-3213(93)90175-O} {\bibfield  {journal}
  {\bibinfo  {journal} {Nucl. Phys. B}\ }\textbf {\bibinfo {volume} {406}},\
  \bibinfo {pages} {423} (\bibinfo {year} {1993})}\BibitemShut {NoStop}%
\bibitem [{\citenamefont {Capozzi}\ and\ \citenamefont
  {Saviano}(2022)}]{capozzi2022neutrino}%
  \BibitemOpen
  \bibfield  {author} {\bibinfo {author} {\bibfnamefont {F.}~\bibnamefont
  {Capozzi}}\ and\ \bibinfo {author} {\bibfnamefont {N.}~\bibnamefont
  {Saviano}},\ }\href {\doibase 10.3390/universe8020094} {\bibfield  {journal}
  {\bibinfo  {journal} {Universe}\ }\textbf {\bibinfo {volume} {8}},\ \bibinfo
  {pages} {94} (\bibinfo {year} {2022})},\ \Eprint
  {http://arxiv.org/abs/2202.02494} {2202.02494} \BibitemShut {NoStop}%
\bibitem [{\citenamefont {Volpe}(2024)}]{volpe2024neutrinos}%
  \BibitemOpen
  \bibfield  {author} {\bibinfo {author} {\bibfnamefont {M.~C.}\ \bibnamefont
  {Volpe}},\ }\href {\doibase 10.1103/RevModPhys.96.025004} {\bibfield
  {journal} {\bibinfo  {journal} {Rev. Mod. Phys.}\ }\textbf {\bibinfo {volume}
  {96}},\ \bibinfo {pages} {025004} (\bibinfo {year} {2024})},\ \Eprint
  {http://arxiv.org/abs/2301.11814} {2301.11814} \BibitemShut {NoStop}%
\bibitem [{\citenamefont {Fischer}\ \emph {et~al.}(2024)\citenamefont
  {Fischer}, \citenamefont {Guo}, \citenamefont {Langanke}, \citenamefont
  {{Martinez-Pinedo}}, \citenamefont {Qian},\ and\ \citenamefont
  {Wu}}]{fischer2024neutrinos}%
  \BibitemOpen
  \bibfield  {author} {\bibinfo {author} {\bibfnamefont {T.}~\bibnamefont
  {Fischer}}, \bibinfo {author} {\bibfnamefont {G.}~\bibnamefont {Guo}},
  \bibinfo {author} {\bibfnamefont {K.}~\bibnamefont {Langanke}}, \bibinfo
  {author} {\bibfnamefont {G.}~\bibnamefont {{Martinez-Pinedo}}}, \bibinfo
  {author} {\bibfnamefont {Y.-Z.}\ \bibnamefont {Qian}}, \ and\ \bibinfo
  {author} {\bibfnamefont {M.-R.}\ \bibnamefont {Wu}},\ }\href {\doibase
  10.1016/j.ppnp.2024.104107} {\bibfield  {journal} {\bibinfo  {journal} {Prog.
  Part. Nucl. Phys.}\ }\textbf {\bibinfo {volume} {137}},\ \bibinfo {pages}
  {104107} (\bibinfo {year} {2024})},\ \Eprint
  {http://arxiv.org/abs/2308.03962} {2308.03962} \BibitemShut {NoStop}%
\bibitem [{\citenamefont {Stapleford}\ \emph {et~al.}(2020)\citenamefont
  {Stapleford}, \citenamefont {Fr{\"o}hlich},\ and\ \citenamefont
  {Kneller}}]{stapleford2020coupling}%
  \BibitemOpen
  \bibfield  {author} {\bibinfo {author} {\bibfnamefont {C.~J.}\ \bibnamefont
  {Stapleford}}, \bibinfo {author} {\bibfnamefont {C.}~\bibnamefont
  {Fr{\"o}hlich}}, \ and\ \bibinfo {author} {\bibfnamefont {J.~P.}\
  \bibnamefont {Kneller}},\ }\href {\doibase 10.1103/PhysRevD.102.081301}
  {\bibfield  {journal} {\bibinfo  {journal} {Phys. Rev. D}\ }\textbf {\bibinfo
  {volume} {102}},\ \bibinfo {pages} {081301} (\bibinfo {year} {2020})},\
  \Eprint {http://arxiv.org/abs/1910.04172} {1910.04172} \BibitemShut {NoStop}%
\bibitem [{\citenamefont {Xiong}\ \emph {et~al.}(2020)\citenamefont {Xiong},
  \citenamefont {Sieverding}, \citenamefont {Sen},\ and\ \citenamefont
  {Qian}}]{xiong2020potential}%
  \BibitemOpen
  \bibfield  {author} {\bibinfo {author} {\bibfnamefont {Z.}~\bibnamefont
  {Xiong}}, \bibinfo {author} {\bibfnamefont {A.}~\bibnamefont {Sieverding}},
  \bibinfo {author} {\bibfnamefont {M.}~\bibnamefont {Sen}}, \ and\ \bibinfo
  {author} {\bibfnamefont {Y.-Z.}\ \bibnamefont {Qian}},\ }\href {\doibase
  10.3847/1538-4357/abac5e} {\bibfield  {journal} {\bibinfo  {journal}
  {Astrophys. J.}\ }\textbf {\bibinfo {volume} {900}},\ \bibinfo {pages} {144}
  (\bibinfo {year} {2020})},\ \Eprint {http://arxiv.org/abs/2006.11414}
  {2006.11414} \BibitemShut {NoStop}%
\bibitem [{\citenamefont {Li}\ and\ \citenamefont
  {Siegel}(2021)}]{li2021neutrino}%
  \BibitemOpen
  \bibfield  {author} {\bibinfo {author} {\bibfnamefont {X.}~\bibnamefont
  {Li}}\ and\ \bibinfo {author} {\bibfnamefont {D.~M.}\ \bibnamefont
  {Siegel}},\ }\href {\doibase 10.1103/PhysRevLett.126.251101} {\bibfield
  {journal} {\bibinfo  {journal} {Phys. Rev. Lett.}\ }\textbf {\bibinfo
  {volume} {126}},\ \bibinfo {pages} {251101} (\bibinfo {year} {2021})},\
  \Eprint {http://arxiv.org/abs/2103.02616} {2103.02616} \BibitemShut {NoStop}%
\bibitem [{\citenamefont {Just}\ \emph {et~al.}(2022)\citenamefont {Just},
  \citenamefont {Abbar}, \citenamefont {Wu}, \citenamefont {Tamborra},
  \citenamefont {Janka},\ and\ \citenamefont {Capozzi}}]{just2022fast}%
  \BibitemOpen
  \bibfield  {author} {\bibinfo {author} {\bibfnamefont {O.}~\bibnamefont
  {Just}}, \bibinfo {author} {\bibfnamefont {S.}~\bibnamefont {Abbar}},
  \bibinfo {author} {\bibfnamefont {M.~R.}\ \bibnamefont {Wu}}, \bibinfo
  {author} {\bibfnamefont {I.}~\bibnamefont {Tamborra}}, \bibinfo {author}
  {\bibfnamefont {H.~T.}\ \bibnamefont {Janka}}, \ and\ \bibinfo {author}
  {\bibfnamefont {F.}~\bibnamefont {Capozzi}},\ }\href {\doibase
  10.1103/PhysRevD.105.083024} {\bibfield  {journal} {\bibinfo  {journal}
  {Phys. Rev. D}\ }\textbf {\bibinfo {volume} {105}},\ \bibinfo {pages}
  {083024} (\bibinfo {year} {2022})},\ \Eprint
  {http://arxiv.org/abs/2203.16559} {2203.16559} \BibitemShut {NoStop}%
\bibitem [{\citenamefont {Fern{\'a}ndez}\ \emph {et~al.}(2022)\citenamefont
  {Fern{\'a}ndez}, \citenamefont {Richers}, \citenamefont {Mulyk},\ and\
  \citenamefont {Fahlman}}]{fernandez2022fast}%
  \BibitemOpen
  \bibfield  {author} {\bibinfo {author} {\bibfnamefont {R.}~\bibnamefont
  {Fern{\'a}ndez}}, \bibinfo {author} {\bibfnamefont {S.}~\bibnamefont
  {Richers}}, \bibinfo {author} {\bibfnamefont {N.}~\bibnamefont {Mulyk}}, \
  and\ \bibinfo {author} {\bibfnamefont {S.}~\bibnamefont {Fahlman}},\ }\href
  {\doibase 10.1103/PhysRevD.106.103003} {\bibfield  {journal} {\bibinfo
  {journal} {Phys. Rev. D}\ }\textbf {\bibinfo {volume} {106}},\ \bibinfo
  {pages} {103003} (\bibinfo {year} {2022})},\ \Eprint
  {http://arxiv.org/abs/2207.10680} {2207.10680} \BibitemShut {NoStop}%
\bibitem [{\citenamefont {Fujimoto}\ and\ \citenamefont
  {Nagakura}(2022)}]{fujimoto2022explosive}%
  \BibitemOpen
  \bibfield  {author} {\bibinfo {author} {\bibfnamefont {S.-i.}\ \bibnamefont
  {Fujimoto}}\ and\ \bibinfo {author} {\bibfnamefont {H.}~\bibnamefont
  {Nagakura}},\ }\href {\doibase 10.1093/mnras/stac3763} {\bibfield  {journal}
  {\bibinfo  {journal} {Mon. Not. Roy. Astron. Soc.}\ }\textbf {\bibinfo
  {volume} {519}},\ \bibinfo {pages} {2623} (\bibinfo {year} {2022})},\ \Eprint
  {http://arxiv.org/abs/2210.02106} {2210.02106} \BibitemShut {NoStop}%
\bibitem [{\citenamefont {Nagakura}(2023{\natexlab{a}})}]{nagakura2023roles}%
  \BibitemOpen
  \bibfield  {author} {\bibinfo {author} {\bibfnamefont {H.}~\bibnamefont
  {Nagakura}},\ }\href {\doibase 10.1103/PhysRevLett.130.211401} {\bibfield
  {journal} {\bibinfo  {journal} {Phys. Rev. Lett.}\ }\textbf {\bibinfo
  {volume} {130}},\ \bibinfo {pages} {211401} (\bibinfo {year}
  {2023}{\natexlab{a}})},\ \Eprint {http://arxiv.org/abs/2301.10785}
  {2301.10785} \BibitemShut {NoStop}%
\bibitem [{\citenamefont {Grohs}\ \emph {et~al.}(2023)\citenamefont {Grohs},
  \citenamefont {Richers}, \citenamefont {Couch}, \citenamefont {Foucart},
  \citenamefont {Kneller},\ and\ \citenamefont
  {McLaughlin}}]{grohs2023neutrino}%
  \BibitemOpen
  \bibfield  {author} {\bibinfo {author} {\bibfnamefont {E.}~\bibnamefont
  {Grohs}}, \bibinfo {author} {\bibfnamefont {S.}~\bibnamefont {Richers}},
  \bibinfo {author} {\bibfnamefont {S.~M.}\ \bibnamefont {Couch}}, \bibinfo
  {author} {\bibfnamefont {F.}~\bibnamefont {Foucart}}, \bibinfo {author}
  {\bibfnamefont {J.~P.}\ \bibnamefont {Kneller}}, \ and\ \bibinfo {author}
  {\bibfnamefont {G.~C.}\ \bibnamefont {McLaughlin}},\ }\href {\doibase
  10.1016/j.physletb.2023.138210} {\bibfield  {journal} {\bibinfo  {journal}
  {Phys. Lett. B}\ }\textbf {\bibinfo {volume} {846}},\ \bibinfo {pages}
  {138210} (\bibinfo {year} {2023})},\ \Eprint
  {http://arxiv.org/abs/2207.02214} {2207.02214} \BibitemShut {NoStop}%
\bibitem [{\citenamefont {Ehring}\ \emph
  {et~al.}(2023{\natexlab{a}})\citenamefont {Ehring}, \citenamefont {Abbar},
  \citenamefont {Janka}, \citenamefont {Raffelt},\ and\ \citenamefont
  {Tamborra}}]{ehring2023fast}%
  \BibitemOpen
  \bibfield  {author} {\bibinfo {author} {\bibfnamefont {J.}~\bibnamefont
  {Ehring}}, \bibinfo {author} {\bibfnamefont {S.}~\bibnamefont {Abbar}},
  \bibinfo {author} {\bibfnamefont {H.~T.}\ \bibnamefont {Janka}}, \bibinfo
  {author} {\bibfnamefont {G.}~\bibnamefont {Raffelt}}, \ and\ \bibinfo
  {author} {\bibfnamefont {I.}~\bibnamefont {Tamborra}},\ }\href {\doibase
  10.1103/PhysRevD.107.103034} {\bibfield  {journal} {\bibinfo  {journal}
  {Phys. Rev. D}\ }\textbf {\bibinfo {volume} {107}},\ \bibinfo {pages}
  {103034} (\bibinfo {year} {2023}{\natexlab{a}})},\ \Eprint
  {http://arxiv.org/abs/2301.11938} {2301.11938} \BibitemShut {NoStop}%
\bibitem [{\citenamefont {Ehring}\ \emph
  {et~al.}(2023{\natexlab{b}})\citenamefont {Ehring}, \citenamefont {Abbar},
  \citenamefont {Janka}, \citenamefont {Raffelt},\ and\ \citenamefont
  {Tamborra}}]{ehring2023fast1}%
  \BibitemOpen
  \bibfield  {author} {\bibinfo {author} {\bibfnamefont {J.}~\bibnamefont
  {Ehring}}, \bibinfo {author} {\bibfnamefont {S.}~\bibnamefont {Abbar}},
  \bibinfo {author} {\bibfnamefont {H.~T.}\ \bibnamefont {Janka}}, \bibinfo
  {author} {\bibfnamefont {G.}~\bibnamefont {Raffelt}}, \ and\ \bibinfo
  {author} {\bibfnamefont {I.}~\bibnamefont {Tamborra}},\ }\href {\doibase
  10.1103/PhysRevLett.131.061401} {\bibfield  {journal} {\bibinfo  {journal}
  {Phys. Rev. Lett.}\ }\textbf {\bibinfo {volume} {131}},\ \bibinfo {pages}
  {061401} (\bibinfo {year} {2023}{\natexlab{b}})},\ \Eprint
  {http://arxiv.org/abs/2305.11207} {2305.11207} \BibitemShut {NoStop}%
\bibitem [{\citenamefont {Xiong}\ \emph {et~al.}(2024)\citenamefont {Xiong},
  \citenamefont {Wu}, \citenamefont {George}, \citenamefont {Lin},
  \citenamefont {Largani}, \citenamefont {Fischer},\ and\ \citenamefont
  {{Mart{\'i}nez-Pinedo}}}]{xiong2024fast}%
  \BibitemOpen
  \bibfield  {author} {\bibinfo {author} {\bibfnamefont {Z.}~\bibnamefont
  {Xiong}}, \bibinfo {author} {\bibfnamefont {M.-R.}\ \bibnamefont {Wu}},
  \bibinfo {author} {\bibfnamefont {M.}~\bibnamefont {George}}, \bibinfo
  {author} {\bibfnamefont {C.-Y.}\ \bibnamefont {Lin}}, \bibinfo {author}
  {\bibfnamefont {N.~K.}\ \bibnamefont {Largani}}, \bibinfo {author}
  {\bibfnamefont {T.}~\bibnamefont {Fischer}}, \ and\ \bibinfo {author}
  {\bibfnamefont {G.}~\bibnamefont {{Mart{\'i}nez-Pinedo}}},\ }\href {\doibase
  10.1103/PhysRevD.109.123008} {\bibfield  {journal} {\bibinfo  {journal}
  {Phys. Rev. D}\ }\textbf {\bibinfo {volume} {109}},\ \bibinfo {pages}
  {123008} (\bibinfo {year} {2024})},\ \Eprint
  {http://arxiv.org/abs/2402.19252} {2402.19252} \BibitemShut {NoStop}%
\bibitem [{\citenamefont {Sawyer}(2009)}]{sawyer2009multiangle}%
  \BibitemOpen
  \bibfield  {author} {\bibinfo {author} {\bibfnamefont {R.~F.}\ \bibnamefont
  {Sawyer}},\ }\href {\doibase 10.1103/PhysRevD.79.105003} {\bibfield
  {journal} {\bibinfo  {journal} {Phys. Rev. D}\ }\textbf {\bibinfo {volume}
  {79}},\ \bibinfo {pages} {105003} (\bibinfo {year} {2009})},\ \Eprint
  {http://arxiv.org/abs/0803.4319} {0803.4319} \BibitemShut {NoStop}%
\bibitem [{\citenamefont {Sawyer}(2016)}]{sawyer2016neutrino}%
  \BibitemOpen
  \bibfield  {author} {\bibinfo {author} {\bibfnamefont {R.~F.}\ \bibnamefont
  {Sawyer}},\ }\href {\doibase 10.1103/PhysRevLett.116.081101} {\bibfield
  {journal} {\bibinfo  {journal} {Phys. Rev. Lett.}\ }\textbf {\bibinfo
  {volume} {116}},\ \bibinfo {pages} {081101} (\bibinfo {year} {2016})},\
  \Eprint {http://arxiv.org/abs/1509.03323} {1509.03323} \BibitemShut {NoStop}%
\bibitem [{\citenamefont {Abbar}\ \emph {et~al.}(2019)\citenamefont {Abbar},
  \citenamefont {Duan}, \citenamefont {Sumiyoshi}, \citenamefont {Takiwaki},\
  and\ \citenamefont {Volpe}}]{abbar2019occurrence}%
  \BibitemOpen
  \bibfield  {author} {\bibinfo {author} {\bibfnamefont {S.}~\bibnamefont
  {Abbar}}, \bibinfo {author} {\bibfnamefont {H.}~\bibnamefont {Duan}},
  \bibinfo {author} {\bibfnamefont {K.}~\bibnamefont {Sumiyoshi}}, \bibinfo
  {author} {\bibfnamefont {T.}~\bibnamefont {Takiwaki}}, \ and\ \bibinfo
  {author} {\bibfnamefont {M.~C.}\ \bibnamefont {Volpe}},\ }\href {\doibase
  10.1103/PhysRevD.100.043004} {\bibfield  {journal} {\bibinfo  {journal}
  {Phys. Rev. D}\ }\textbf {\bibinfo {volume} {100}},\ \bibinfo {pages}
  {043004} (\bibinfo {year} {2019})},\ \Eprint
  {http://arxiv.org/abs/1812.06883} {1812.06883} \BibitemShut {NoStop}%
\bibitem [{\citenamefont {Delfan~Azari}\ \emph {et~al.}(2019)\citenamefont
  {Delfan~Azari}, \citenamefont {Yamada}, \citenamefont {Morinaga},
  \citenamefont {Iwakami}, \citenamefont {Okawa}, \citenamefont {Nagakura},\
  and\ \citenamefont {Sumiyoshi}}]{delfanazari2019linear}%
  \BibitemOpen
  \bibfield  {author} {\bibinfo {author} {\bibfnamefont {M.}~\bibnamefont
  {Delfan~Azari}}, \bibinfo {author} {\bibfnamefont {S.}~\bibnamefont
  {Yamada}}, \bibinfo {author} {\bibfnamefont {T.}~\bibnamefont {Morinaga}},
  \bibinfo {author} {\bibfnamefont {W.}~\bibnamefont {Iwakami}}, \bibinfo
  {author} {\bibfnamefont {H.}~\bibnamefont {Okawa}}, \bibinfo {author}
  {\bibfnamefont {H.}~\bibnamefont {Nagakura}}, \ and\ \bibinfo {author}
  {\bibfnamefont {K.}~\bibnamefont {Sumiyoshi}},\ }\href {\doibase
  10.1103/PhysRevD.99.103011} {\bibfield  {journal} {\bibinfo  {journal} {Phys.
  Rev. D}\ }\textbf {\bibinfo {volume} {99}},\ \bibinfo {pages} {103011}
  (\bibinfo {year} {2019})},\ \Eprint {http://arxiv.org/abs/1902.07467}
  {1902.07467} \BibitemShut {NoStop}%
\bibitem [{\citenamefont {Delfan~Azari}\ \emph {et~al.}(2020)\citenamefont
  {Delfan~Azari}, \citenamefont {Yamada}, \citenamefont {Morinaga},
  \citenamefont {Nagakura}, \citenamefont {Furusawa}, \citenamefont {Harada},
  \citenamefont {Okawa}, \citenamefont {Iwakami},\ and\ \citenamefont
  {Sumiyoshi}}]{delfanazari2020fast}%
  \BibitemOpen
  \bibfield  {author} {\bibinfo {author} {\bibfnamefont {M.}~\bibnamefont
  {Delfan~Azari}}, \bibinfo {author} {\bibfnamefont {S.}~\bibnamefont
  {Yamada}}, \bibinfo {author} {\bibfnamefont {T.}~\bibnamefont {Morinaga}},
  \bibinfo {author} {\bibfnamefont {H.}~\bibnamefont {Nagakura}}, \bibinfo
  {author} {\bibfnamefont {S.}~\bibnamefont {Furusawa}}, \bibinfo {author}
  {\bibfnamefont {A.}~\bibnamefont {Harada}}, \bibinfo {author} {\bibfnamefont
  {H.}~\bibnamefont {Okawa}}, \bibinfo {author} {\bibfnamefont
  {W.}~\bibnamefont {Iwakami}}, \ and\ \bibinfo {author} {\bibfnamefont
  {K.}~\bibnamefont {Sumiyoshi}},\ }\href {\doibase
  10.1103/PhysRevD.101.023018} {\bibfield  {journal} {\bibinfo  {journal}
  {Phys. Rev. D}\ }\textbf {\bibinfo {volume} {101}},\ \bibinfo {pages}
  {023018} (\bibinfo {year} {2020})},\ \Eprint
  {http://arxiv.org/abs/1910.06176} {1910.06176} \BibitemShut {NoStop}%
\bibitem [{\citenamefont {Nagakura}\ \emph {et~al.}(2019)\citenamefont
  {Nagakura}, \citenamefont {Morinaga}, \citenamefont {Kato},\ and\
  \citenamefont {Yamada}}]{nagakura2019fastpairwise}%
  \BibitemOpen
  \bibfield  {author} {\bibinfo {author} {\bibfnamefont {H.}~\bibnamefont
  {Nagakura}}, \bibinfo {author} {\bibfnamefont {T.}~\bibnamefont {Morinaga}},
  \bibinfo {author} {\bibfnamefont {C.}~\bibnamefont {Kato}}, \ and\ \bibinfo
  {author} {\bibfnamefont {S.}~\bibnamefont {Yamada}},\ }\href {\doibase
  10.3847/1538-4357/ab4cf2} {\bibfield  {journal} {\bibinfo  {journal} {The
  Astrophysical Journal}\ }\textbf {\bibinfo {volume} {886}},\ \bibinfo {pages}
  {139} (\bibinfo {year} {2019})},\ \Eprint {http://arxiv.org/abs/1910.04288}
  {1910.04288} \BibitemShut {NoStop}%
\bibitem [{\citenamefont {Abbar}\ \emph {et~al.}(2020)\citenamefont {Abbar},
  \citenamefont {Duan}, \citenamefont {Sumiyoshi}, \citenamefont {Takiwaki},\
  and\ \citenamefont {Volpe}}]{abbar2020fast}%
  \BibitemOpen
  \bibfield  {author} {\bibinfo {author} {\bibfnamefont {S.}~\bibnamefont
  {Abbar}}, \bibinfo {author} {\bibfnamefont {H.}~\bibnamefont {Duan}},
  \bibinfo {author} {\bibfnamefont {K.}~\bibnamefont {Sumiyoshi}}, \bibinfo
  {author} {\bibfnamefont {T.}~\bibnamefont {Takiwaki}}, \ and\ \bibinfo
  {author} {\bibfnamefont {M.~C.}\ \bibnamefont {Volpe}},\ }\href {\doibase
  10.1103/PhysRevD.101.043016} {\bibfield  {journal} {\bibinfo  {journal}
  {Phys. Rev. D}\ }\textbf {\bibinfo {volume} {101}},\ \bibinfo {pages}
  {043016} (\bibinfo {year} {2020})},\ \Eprint
  {http://arxiv.org/abs/1911.01983} {1911.01983} \BibitemShut {NoStop}%
\bibitem [{\citenamefont {Glas}\ \emph {et~al.}(2020)\citenamefont {Glas},
  \citenamefont {Janka}, \citenamefont {Capozzi}, \citenamefont {Sen},
  \citenamefont {Dasgupta}, \citenamefont {Mirizzi},\ and\ \citenamefont
  {Sigl}}]{glas2020fast}%
  \BibitemOpen
  \bibfield  {author} {\bibinfo {author} {\bibfnamefont {R.}~\bibnamefont
  {Glas}}, \bibinfo {author} {\bibfnamefont {H.-T.}\ \bibnamefont {Janka}},
  \bibinfo {author} {\bibfnamefont {F.}~\bibnamefont {Capozzi}}, \bibinfo
  {author} {\bibfnamefont {M.}~\bibnamefont {Sen}}, \bibinfo {author}
  {\bibfnamefont {B.}~\bibnamefont {Dasgupta}}, \bibinfo {author}
  {\bibfnamefont {A.}~\bibnamefont {Mirizzi}}, \ and\ \bibinfo {author}
  {\bibfnamefont {G.}~\bibnamefont {Sigl}},\ }\href {\doibase
  10.1103/PhysRevD.101.063001} {\bibfield  {journal} {\bibinfo  {journal}
  {Phys. Rev. D}\ }\textbf {\bibinfo {volume} {101}},\ \bibinfo {pages}
  {063001} (\bibinfo {year} {2020})},\ \Eprint
  {http://arxiv.org/abs/1912.00274} {1912.00274} \BibitemShut {NoStop}%
\bibitem [{\citenamefont {Nagakura}\ \emph {et~al.}(2021)\citenamefont
  {Nagakura}, \citenamefont {Burrows}, \citenamefont {Johns},\ and\
  \citenamefont {Fuller}}]{nagakura2021where}%
  \BibitemOpen
  \bibfield  {author} {\bibinfo {author} {\bibfnamefont {H.}~\bibnamefont
  {Nagakura}}, \bibinfo {author} {\bibfnamefont {A.}~\bibnamefont {Burrows}},
  \bibinfo {author} {\bibfnamefont {L.}~\bibnamefont {Johns}}, \ and\ \bibinfo
  {author} {\bibfnamefont {G.~M.}\ \bibnamefont {Fuller}},\ }\href {\doibase
  10.1103/PhysRevD.104.083025} {\bibfield  {journal} {\bibinfo  {journal}
  {Phys. Rev. D}\ }\textbf {\bibinfo {volume} {104}},\ \bibinfo {pages}
  {083025} (\bibinfo {year} {2021})},\ \Eprint
  {http://arxiv.org/abs/2108.07281} {2108.07281} \BibitemShut {NoStop}%
\bibitem [{\citenamefont {Abbar}\ \emph {et~al.}(2021)\citenamefont {Abbar},
  \citenamefont {Capozzi}, \citenamefont {Glas}, \citenamefont {Janka},\ and\
  \citenamefont {Tamborra}}]{abbar2021characteristics}%
  \BibitemOpen
  \bibfield  {author} {\bibinfo {author} {\bibfnamefont {S.}~\bibnamefont
  {Abbar}}, \bibinfo {author} {\bibfnamefont {F.}~\bibnamefont {Capozzi}},
  \bibinfo {author} {\bibfnamefont {R.}~\bibnamefont {Glas}}, \bibinfo {author}
  {\bibfnamefont {H.~T.}\ \bibnamefont {Janka}}, \ and\ \bibinfo {author}
  {\bibfnamefont {I.}~\bibnamefont {Tamborra}},\ }\href {\doibase
  10.1103/PhysRevD.103.063033} {\bibfield  {journal} {\bibinfo  {journal}
  {Phys. Rev. D}\ }\textbf {\bibinfo {volume} {103}},\ \bibinfo {pages}
  {063033} (\bibinfo {year} {2021})},\ \Eprint
  {http://arxiv.org/abs/2012.06594} {2012.06594} \BibitemShut {NoStop}%
\bibitem [{\citenamefont {Harada}\ and\ \citenamefont
  {Nagakura}(2022)}]{harada2022prospects}%
  \BibitemOpen
  \bibfield  {author} {\bibinfo {author} {\bibfnamefont {A.}~\bibnamefont
  {Harada}}\ and\ \bibinfo {author} {\bibfnamefont {H.}~\bibnamefont
  {Nagakura}},\ }\href {\doibase 10.3847/1538-4357/ac38a0} {\bibfield
  {journal} {\bibinfo  {journal} {Astrophys. J.}\ }\textbf {\bibinfo {volume}
  {924}},\ \bibinfo {pages} {109} (\bibinfo {year} {2022})},\ \Eprint
  {http://arxiv.org/abs/2110.08291} {2110.08291} \BibitemShut {NoStop}%
\bibitem [{\citenamefont {Wu}\ and\ \citenamefont
  {Tamborra}(2017)}]{wu2017fast}%
  \BibitemOpen
  \bibfield  {author} {\bibinfo {author} {\bibfnamefont {M.~R.}\ \bibnamefont
  {Wu}}\ and\ \bibinfo {author} {\bibfnamefont {I.}~\bibnamefont {Tamborra}},\
  }\href {\doibase 10.1103/PhysRevD.95.103007} {\bibfield  {journal} {\bibinfo
  {journal} {Phys. Rev. D}\ }\textbf {\bibinfo {volume} {95}},\ \bibinfo
  {pages} {103007} (\bibinfo {year} {2017})},\ \Eprint
  {http://arxiv.org/abs/1701.06580} {1701.06580} \BibitemShut {NoStop}%
\bibitem [{\citenamefont {Wu}\ \emph {et~al.}(2017)\citenamefont {Wu},
  \citenamefont {Tamborra}, \citenamefont {Just},\ and\ \citenamefont
  {Janka}}]{wu2017imprints}%
  \BibitemOpen
  \bibfield  {author} {\bibinfo {author} {\bibfnamefont {M.~R.}\ \bibnamefont
  {Wu}}, \bibinfo {author} {\bibfnamefont {I.}~\bibnamefont {Tamborra}},
  \bibinfo {author} {\bibfnamefont {O.}~\bibnamefont {Just}}, \ and\ \bibinfo
  {author} {\bibfnamefont {H.~T.}\ \bibnamefont {Janka}},\ }\href {\doibase
  10.1103/PhysRevD.96.123015} {\bibfield  {journal} {\bibinfo  {journal} {Phys.
  Rev. D}\ }\textbf {\bibinfo {volume} {96}},\ \bibinfo {pages} {123015}
  (\bibinfo {year} {2017})},\ \Eprint {http://arxiv.org/abs/1711.00477}
  {1711.00477} \BibitemShut {NoStop}%
\bibitem [{\citenamefont {George}\ \emph {et~al.}(2020)\citenamefont {George},
  \citenamefont {Wu}, \citenamefont {Tamborra}, \citenamefont
  {{Ardevol-Pulpillo}},\ and\ \citenamefont {Janka}}]{george2020fast}%
  \BibitemOpen
  \bibfield  {author} {\bibinfo {author} {\bibfnamefont {M.}~\bibnamefont
  {George}}, \bibinfo {author} {\bibfnamefont {M.-R.}\ \bibnamefont {Wu}},
  \bibinfo {author} {\bibfnamefont {I.}~\bibnamefont {Tamborra}}, \bibinfo
  {author} {\bibfnamefont {R.}~\bibnamefont {{Ardevol-Pulpillo}}}, \ and\
  \bibinfo {author} {\bibfnamefont {H.-T.}\ \bibnamefont {Janka}},\ }\href
  {\doibase 10.1103/PhysRevD.102.103015} {\bibfield  {journal} {\bibinfo
  {journal} {Phys. Rev. D}\ }\textbf {\bibinfo {volume} {102}},\ \bibinfo
  {pages} {103015} (\bibinfo {year} {2020})},\ \Eprint
  {http://arxiv.org/abs/2009.04046} {2009.04046} \BibitemShut {NoStop}%
\bibitem [{\citenamefont {Richers}(2022)}]{richers2022evaluating}%
  \BibitemOpen
  \bibfield  {author} {\bibinfo {author} {\bibfnamefont {S.}~\bibnamefont
  {Richers}},\ }\href {\doibase 10.1103/PhysRevD.106.083005} {\bibfield
  {journal} {\bibinfo  {journal} {Phys. Rev. D}\ }\textbf {\bibinfo {volume}
  {106}},\ \bibinfo {pages} {083005} (\bibinfo {year} {2022})},\ \Eprint
  {http://arxiv.org/abs/2206.08444} {2206.08444} \BibitemShut {NoStop}%
\bibitem [{\citenamefont {Mukhopadhyay}\ \emph {et~al.}(2024)\citenamefont
  {Mukhopadhyay}, \citenamefont {Miller},\ and\ \citenamefont
  {McLaughlin}}]{mukhopadhyay2024time}%
  \BibitemOpen
  \bibfield  {author} {\bibinfo {author} {\bibfnamefont {P.}~\bibnamefont
  {Mukhopadhyay}}, \bibinfo {author} {\bibfnamefont {J.}~\bibnamefont
  {Miller}}, \ and\ \bibinfo {author} {\bibfnamefont {G.~C.}\ \bibnamefont
  {McLaughlin}},\ }\href {\doibase 10.3847/1538-4357/ad6c42} {\bibfield
  {journal} {\bibinfo  {journal} {Astrophys. J.}\ }\textbf {\bibinfo {volume}
  {974}},\ \bibinfo {pages} {110} (\bibinfo {year} {2024})},\ \Eprint
  {http://arxiv.org/abs/2404.17938} {arXiv:2404.17938 [astro-ph.HE]}
  \BibitemShut {NoStop}%
\bibitem [{\citenamefont {Tamborra}\ and\ \citenamefont
  {Shalgar}(2021)}]{tamborra2021new}%
  \BibitemOpen
  \bibfield  {author} {\bibinfo {author} {\bibfnamefont {I.}~\bibnamefont
  {Tamborra}}\ and\ \bibinfo {author} {\bibfnamefont {S.}~\bibnamefont
  {Shalgar}},\ }\href {\doibase 10.1146/annurev-nucl-102920-050505} {\bibfield
  {journal} {\bibinfo  {journal} {Ann. Rev. Nucl. Part. Sci.}\ }\textbf
  {\bibinfo {volume} {71}},\ \bibinfo {pages} {165} (\bibinfo {year} {2021})},\
  \Eprint {http://arxiv.org/abs/2011.01948} {2011.01948} \BibitemShut {NoStop}%
\bibitem [{\citenamefont {Richers}\ and\ \citenamefont
  {Sen}(2023)}]{richers2022fast}%
  \BibitemOpen
  \bibfield  {author} {\bibinfo {author} {\bibfnamefont {S.}~\bibnamefont
  {Richers}}\ and\ \bibinfo {author} {\bibfnamefont {M.}~\bibnamefont {Sen}},\
  }\href {\doibase 10.1007/978-981-19-6345-2_125} {\bibfield  {journal}
  {\bibinfo  {journal} {Handbook of Nuclear Physics}\ ,\ \bibinfo {pages}
  {3771}} (\bibinfo {year} {2023})},\ \Eprint {http://arxiv.org/abs/2207.03561}
  {2207.03561} \BibitemShut {NoStop}%
\bibitem [{\citenamefont {Vlasenko}\ \emph {et~al.}(2014)\citenamefont
  {Vlasenko}, \citenamefont {Fuller},\ and\ \citenamefont
  {Cirigliano}}]{vlasenko2014neutrino}%
  \BibitemOpen
  \bibfield  {author} {\bibinfo {author} {\bibfnamefont {A.}~\bibnamefont
  {Vlasenko}}, \bibinfo {author} {\bibfnamefont {G.~M.}\ \bibnamefont
  {Fuller}}, \ and\ \bibinfo {author} {\bibfnamefont {V.}~\bibnamefont
  {Cirigliano}},\ }\href {\doibase 10.1103/PhysRevD.89.105004} {\bibfield
  {journal} {\bibinfo  {journal} {Phys. Rev. D}\ }\textbf {\bibinfo {volume}
  {89}},\ \bibinfo {pages} {105004} (\bibinfo {year} {2014})},\ \Eprint
  {http://arxiv.org/abs/1309.2628} {1309.2628} \BibitemShut {NoStop}%
\bibitem [{\citenamefont {Volpe}(2015)}]{volpe2015neutrino}%
  \BibitemOpen
  \bibfield  {author} {\bibinfo {author} {\bibfnamefont {C.}~\bibnamefont
  {Volpe}},\ }\href {\doibase 10.1142/S0218301315410098} {\bibfield  {journal}
  {\bibinfo  {journal} {Int. J. Mod. Phys. E}\ }\textbf {\bibinfo {volume}
  {24}},\ \bibinfo {pages} {1541009} (\bibinfo {year} {2015})},\ \Eprint
  {http://arxiv.org/abs/1506.06222} {1506.06222} \BibitemShut {NoStop}%
\bibitem [{\citenamefont {Bhattacharyya}\ and\ \citenamefont
  {Dasgupta}(2020)}]{bhattacharyya2020latetime}%
  \BibitemOpen
  \bibfield  {author} {\bibinfo {author} {\bibfnamefont {S.}~\bibnamefont
  {Bhattacharyya}}\ and\ \bibinfo {author} {\bibfnamefont {B.}~\bibnamefont
  {Dasgupta}},\ }\href {\doibase 10.1103/PhysRevD.102.063018} {\bibfield
  {journal} {\bibinfo  {journal} {Phys. Rev. D}\ }\textbf {\bibinfo {volume}
  {102}},\ \bibinfo {pages} {063018} (\bibinfo {year} {2020})},\ \Eprint
  {http://arxiv.org/abs/2005.00459} {2005.00459} \BibitemShut {NoStop}%
\bibitem [{\citenamefont {Bhattacharyya}\ and\ \citenamefont
  {Dasgupta}(2021)}]{bhattacharyya2021fast}%
  \BibitemOpen
  \bibfield  {author} {\bibinfo {author} {\bibfnamefont {S.}~\bibnamefont
  {Bhattacharyya}}\ and\ \bibinfo {author} {\bibfnamefont {B.}~\bibnamefont
  {Dasgupta}},\ }\href {\doibase 10.1103/PhysRevLett.126.061302} {\bibfield
  {journal} {\bibinfo  {journal} {Phys. Rev. Lett.}\ }\textbf {\bibinfo
  {volume} {126}},\ \bibinfo {pages} {061302} (\bibinfo {year} {2021})},\
  \Eprint {http://arxiv.org/abs/2009.03337} {2009.03337} \BibitemShut {NoStop}%
\bibitem [{\citenamefont {Wu}\ \emph {et~al.}(2021)\citenamefont {Wu},
  \citenamefont {George}, \citenamefont {Lin},\ and\ \citenamefont
  {Xiong}}]{wu2021collective}%
  \BibitemOpen
  \bibfield  {author} {\bibinfo {author} {\bibfnamefont {M.-R.}\ \bibnamefont
  {Wu}}, \bibinfo {author} {\bibfnamefont {M.}~\bibnamefont {George}}, \bibinfo
  {author} {\bibfnamefont {C.-Y.}\ \bibnamefont {Lin}}, \ and\ \bibinfo
  {author} {\bibfnamefont {Z.}~\bibnamefont {Xiong}},\ }\href {\doibase
  10.1103/PhysRevD.104.103003} {\bibfield  {journal} {\bibinfo  {journal}
  {Phys. Rev. D}\ }\textbf {\bibinfo {volume} {104}},\ \bibinfo {pages}
  {103003} (\bibinfo {year} {2021})},\ \Eprint
  {http://arxiv.org/abs/2108.09886} {2108.09886} \BibitemShut {NoStop}%
\bibitem [{\citenamefont {Richers}\ \emph
  {et~al.}(2021{\natexlab{a}})\citenamefont {Richers}, \citenamefont {Willcox},
  \citenamefont {Ford},\ and\ \citenamefont
  {Myers}}]{richers2021particleincell}%
  \BibitemOpen
  \bibfield  {author} {\bibinfo {author} {\bibfnamefont {S.}~\bibnamefont
  {Richers}}, \bibinfo {author} {\bibfnamefont {D.~E.}\ \bibnamefont
  {Willcox}}, \bibinfo {author} {\bibfnamefont {N.~M.}\ \bibnamefont {Ford}}, \
  and\ \bibinfo {author} {\bibfnamefont {A.}~\bibnamefont {Myers}},\ }\href
  {\doibase 10.1103/PhysRevD.103.083013} {\bibfield  {journal} {\bibinfo
  {journal} {Phys. Rev. D}\ }\textbf {\bibinfo {volume} {103}},\ \bibinfo
  {pages} {083013} (\bibinfo {year} {2021}{\natexlab{a}})},\ \Eprint
  {http://arxiv.org/abs/2101.02745} {2101.02745} \BibitemShut {NoStop}%
\bibitem [{\citenamefont {Richers}\ \emph {et~al.}(2022)\citenamefont
  {Richers}, \citenamefont {Duan}, \citenamefont {Wu}, \citenamefont
  {Bhattacharyya}, \citenamefont {Zaizen}, \citenamefont {George},
  \citenamefont {Lin},\ and\ \citenamefont {Xiong}}]{richers2022code}%
  \BibitemOpen
  \bibfield  {author} {\bibinfo {author} {\bibfnamefont {S.}~\bibnamefont
  {Richers}}, \bibinfo {author} {\bibfnamefont {H.}~\bibnamefont {Duan}},
  \bibinfo {author} {\bibfnamefont {M.-R.}\ \bibnamefont {Wu}}, \bibinfo
  {author} {\bibfnamefont {S.}~\bibnamefont {Bhattacharyya}}, \bibinfo {author}
  {\bibfnamefont {M.}~\bibnamefont {Zaizen}}, \bibinfo {author} {\bibfnamefont
  {M.}~\bibnamefont {George}}, \bibinfo {author} {\bibfnamefont {C.-Y.}\
  \bibnamefont {Lin}}, \ and\ \bibinfo {author} {\bibfnamefont
  {Z.}~\bibnamefont {Xiong}},\ }\href {\doibase 10.1103/PhysRevD.106.043011}
  {\bibfield  {journal} {\bibinfo  {journal} {Phys. Rev. D}\ }\textbf {\bibinfo
  {volume} {106}},\ \bibinfo {pages} {043011} (\bibinfo {year} {2022})},\
  \Eprint {http://arxiv.org/abs/2205.06282} {2205.06282} \BibitemShut {NoStop}%
\bibitem [{\citenamefont {G~Fiorillo}\ and\ \citenamefont
  {Raffelt}(2024)}]{gfiorillo2024fast}%
  \BibitemOpen
  \bibfield  {author} {\bibinfo {author} {\bibfnamefont {D.~F.}\ \bibnamefont
  {G~Fiorillo}}\ and\ \bibinfo {author} {\bibfnamefont {G.~G.}\ \bibnamefont
  {Raffelt}},\ }\href@noop {} {\  (\bibinfo {year} {2024})},\ \Eprint
  {http://arxiv.org/abs/2403.12189} {2403.12189} \BibitemShut {NoStop}%
\bibitem [{\citenamefont {Cornelius}\ \emph
  {et~al.}(2024{\natexlab{a}})\citenamefont {Cornelius}, \citenamefont
  {Shalgar},\ and\ \citenamefont {Tamborra}}]{cornelius2024perturbing}%
  \BibitemOpen
  \bibfield  {author} {\bibinfo {author} {\bibfnamefont {M.}~\bibnamefont
  {Cornelius}}, \bibinfo {author} {\bibfnamefont {S.}~\bibnamefont {Shalgar}},
  \ and\ \bibinfo {author} {\bibfnamefont {I.}~\bibnamefont {Tamborra}},\
  }\href {\doibase 10.1088/1475-7516/2024/02/038} {\bibfield  {journal}
  {\bibinfo  {journal} {JCAP}\ }\textbf {\bibinfo {volume} {02}},\ \bibinfo
  {pages} {038} (\bibinfo {year} {2024}{\natexlab{a}})},\ \Eprint
  {http://arxiv.org/abs/2312.03839} {2312.03839} \BibitemShut {NoStop}%
\bibitem [{\citenamefont {Delfan~Azari}\ \emph {et~al.}(2024)\citenamefont
  {Delfan~Azari}, \citenamefont {Sasaki}, \citenamefont {Takiwaki},\ and\
  \citenamefont {Okawa}}]{azari2024systematic}%
  \BibitemOpen
  \bibfield  {author} {\bibinfo {author} {\bibfnamefont {M.}~\bibnamefont
  {Delfan~Azari}}, \bibinfo {author} {\bibfnamefont {H.}~\bibnamefont
  {Sasaki}}, \bibinfo {author} {\bibfnamefont {T.}~\bibnamefont {Takiwaki}}, \
  and\ \bibinfo {author} {\bibfnamefont {H.}~\bibnamefont {Okawa}},\ }\href
  {\doibase 10.1093/ptep/ptae144} {\bibfield  {journal} {\bibinfo  {journal}
  {PTEP}\ }\textbf {\bibinfo {volume} {2024}},\ \bibinfo {pages} {103E01}
  (\bibinfo {year} {2024})},\ \Eprint {http://arxiv.org/abs/2402.04741}
  {arXiv:2402.04741 [hep-ph]} \BibitemShut {NoStop}%
\bibitem [{\citenamefont {Zaizen}\ and\ \citenamefont
  {Nagakura}(2023)}]{zaizen2023simple}%
  \BibitemOpen
  \bibfield  {author} {\bibinfo {author} {\bibfnamefont {M.}~\bibnamefont
  {Zaizen}}\ and\ \bibinfo {author} {\bibfnamefont {H.}~\bibnamefont
  {Nagakura}},\ }\href {\doibase 10.1103/PhysRevD.107.103022} {\bibfield
  {journal} {\bibinfo  {journal} {Phys. Rev. D}\ }\textbf {\bibinfo {volume}
  {107}},\ \bibinfo {pages} {103022} (\bibinfo {year} {2023})},\ \Eprint
  {http://arxiv.org/abs/2211.09343} {2211.09343} \BibitemShut {NoStop}%
\bibitem [{\citenamefont {Xiong}\ \emph
  {et~al.}(2023{\natexlab{a}})\citenamefont {Xiong}, \citenamefont {Wu},
  \citenamefont {Abbar}, \citenamefont {Bhattacharyya}, \citenamefont
  {George},\ and\ \citenamefont {Lin}}]{xiong2023evaluating}%
  \BibitemOpen
  \bibfield  {author} {\bibinfo {author} {\bibfnamefont {Z.}~\bibnamefont
  {Xiong}}, \bibinfo {author} {\bibfnamefont {R.}~\bibnamefont {Wu}}, \bibinfo
  {author} {\bibfnamefont {S.}~\bibnamefont {Abbar}}, \bibinfo {author}
  {\bibfnamefont {S.}~\bibnamefont {Bhattacharyya}}, \bibinfo {author}
  {\bibfnamefont {M.}~\bibnamefont {George}}, \ and\ \bibinfo {author}
  {\bibfnamefont {C.-Y.}\ \bibnamefont {Lin}},\ }\href {\doibase
  10.1103/PhysRevD.108.063003} {\bibfield  {journal} {\bibinfo  {journal}
  {Phys. Rev. D}\ }\textbf {\bibinfo {volume} {108}},\ \bibinfo {pages}
  {063003} (\bibinfo {year} {2023}{\natexlab{a}})},\ \Eprint
  {http://arxiv.org/abs/2307.11129} {2307.11129} \BibitemShut {NoStop}%
\bibitem [{\citenamefont {{Xiong}}\ \emph {et~al.}(2024)\citenamefont
  {{Xiong}}, \citenamefont {{Wu}}, \citenamefont {{George}},\ and\
  \citenamefont {{Lin}}}]{xiong2024robust}%
  \BibitemOpen
  \bibfield  {author} {\bibinfo {author} {\bibfnamefont {Z.}~\bibnamefont
  {{Xiong}}}, \bibinfo {author} {\bibfnamefont {M.-R.}\ \bibnamefont {{Wu}}},
  \bibinfo {author} {\bibfnamefont {M.}~\bibnamefont {{George}}}, \ and\
  \bibinfo {author} {\bibfnamefont {C.-Y.}\ \bibnamefont {{Lin}}},\ }\href@noop
  {} {\enquote {\bibinfo {title} {{Robust integration of fast flavor
  conversions in classical neutrino transport}},}\ } (\bibinfo {year} {2024}),\
  \Eprint {http://arxiv.org/abs/2403.17269} {2403.17269} \BibitemShut {NoStop}%
\bibitem [{\citenamefont {Capozzi}\ \emph {et~al.}(2019)\citenamefont
  {Capozzi}, \citenamefont {Dasgupta}, \citenamefont {Mirizzi}, \citenamefont
  {Sen},\ and\ \citenamefont {Sigl}}]{capozzi2019collisional}%
  \BibitemOpen
  \bibfield  {author} {\bibinfo {author} {\bibfnamefont {F.}~\bibnamefont
  {Capozzi}}, \bibinfo {author} {\bibfnamefont {B.}~\bibnamefont {Dasgupta}},
  \bibinfo {author} {\bibfnamefont {A.}~\bibnamefont {Mirizzi}}, \bibinfo
  {author} {\bibfnamefont {M.}~\bibnamefont {Sen}}, \ and\ \bibinfo {author}
  {\bibfnamefont {G.}~\bibnamefont {Sigl}},\ }\href {\doibase
  10.1103/PhysRevLett.122.091101} {\bibfield  {journal} {\bibinfo  {journal}
  {Phys. Rev. Lett.}\ }\textbf {\bibinfo {volume} {122}},\ \bibinfo {pages}
  {091101} (\bibinfo {year} {2019})},\ \Eprint
  {http://arxiv.org/abs/1808.06618} {1808.06618} \BibitemShut {NoStop}%
\bibitem [{\citenamefont {{Padilla-Gay}}\ \emph {et~al.}(2021)\citenamefont
  {{Padilla-Gay}}, \citenamefont {Shalgar},\ and\ \citenamefont
  {Tamborra}}]{padilla2021multidimensional}%
  \BibitemOpen
  \bibfield  {author} {\bibinfo {author} {\bibfnamefont {I.}~\bibnamefont
  {{Padilla-Gay}}}, \bibinfo {author} {\bibfnamefont {S.}~\bibnamefont
  {Shalgar}}, \ and\ \bibinfo {author} {\bibfnamefont {I.}~\bibnamefont
  {Tamborra}},\ }\href {\doibase 10.1088/1475-7516/2021/01/017} {\bibfield
  {journal} {\bibinfo  {journal} {JCAP}\ }\textbf {\bibinfo {volume} {01}},\
  \bibinfo {pages} {017} (\bibinfo {year} {2021})},\ \Eprint
  {http://arxiv.org/abs/2009.01843} {2009.01843} \BibitemShut {NoStop}%
\bibitem [{\citenamefont {Nagakura}\ and\ \citenamefont
  {Zaizen}(2022)}]{nagakura2022timedependent}%
  \BibitemOpen
  \bibfield  {author} {\bibinfo {author} {\bibfnamefont {H.}~\bibnamefont
  {Nagakura}}\ and\ \bibinfo {author} {\bibfnamefont {M.}~\bibnamefont
  {Zaizen}},\ }\href {\doibase 10.1103/PhysRevLett.129.261101} {\bibfield
  {journal} {\bibinfo  {journal} {Phys. Rev. Lett.}\ }\textbf {\bibinfo
  {volume} {129}},\ \bibinfo {pages} {261101} (\bibinfo {year} {2022})},\
  \Eprint {http://arxiv.org/abs/2206.04097} {2206.04097} \BibitemShut {NoStop}%
\bibitem [{\citenamefont {Xiong}\ \emph
  {et~al.}(2023{\natexlab{b}})\citenamefont {Xiong}, \citenamefont {Wu},
  \citenamefont {{Mart{\'i}nez-Pinedo}}, \citenamefont {Fischer}, \citenamefont
  {George}, \citenamefont {Lin},\ and\ \citenamefont
  {Johns}}]{xiong2023evolution}%
  \BibitemOpen
  \bibfield  {author} {\bibinfo {author} {\bibfnamefont {Z.}~\bibnamefont
  {Xiong}}, \bibinfo {author} {\bibfnamefont {M.-R.}\ \bibnamefont {Wu}},
  \bibinfo {author} {\bibfnamefont {G.}~\bibnamefont {{Mart{\'i}nez-Pinedo}}},
  \bibinfo {author} {\bibfnamefont {T.}~\bibnamefont {Fischer}}, \bibinfo
  {author} {\bibfnamefont {M.}~\bibnamefont {George}}, \bibinfo {author}
  {\bibfnamefont {C.-Y.}\ \bibnamefont {Lin}}, \ and\ \bibinfo {author}
  {\bibfnamefont {L.}~\bibnamefont {Johns}},\ }\href {\doibase
  10.1103/PhysRevD.107.083016} {\bibfield  {journal} {\bibinfo  {journal}
  {Phys. Rev. D}\ }\textbf {\bibinfo {volume} {107}},\ \bibinfo {pages}
  {083016} (\bibinfo {year} {2023}{\natexlab{b}})},\ \Eprint
  {http://arxiv.org/abs/2210.08254} {2210.08254} \BibitemShut {NoStop}%
\bibitem [{\citenamefont {Nagakura}\ and\ \citenamefont
  {Zaizen}(2023)}]{nagakura2023basic}%
  \BibitemOpen
  \bibfield  {author} {\bibinfo {author} {\bibfnamefont {H.}~\bibnamefont
  {Nagakura}}\ and\ \bibinfo {author} {\bibfnamefont {M.}~\bibnamefont
  {Zaizen}},\ }\href {\doibase 10.1103/PhysRevD.108.123003} {\bibfield
  {journal} {\bibinfo  {journal} {Phys. Rev. D}\ }\textbf {\bibinfo {volume}
  {108}},\ \bibinfo {pages} {123003} (\bibinfo {year} {2023})},\ \Eprint
  {http://arxiv.org/abs/2308.14800} {2308.14800} \BibitemShut {NoStop}%
\bibitem [{\citenamefont {Nagakura}(2023{\natexlab{b}})}]{nagakura2023global}%
  \BibitemOpen
  \bibfield  {author} {\bibinfo {author} {\bibfnamefont {H.}~\bibnamefont
  {Nagakura}},\ }\href {\doibase 10.1103/PhysRevD.108.103014} {\bibfield
  {journal} {\bibinfo  {journal} {Phys. Rev. D}\ }\textbf {\bibinfo {volume}
  {108}},\ \bibinfo {pages} {103014} (\bibinfo {year} {2023}{\natexlab{b}})},\
  \Eprint {http://arxiv.org/abs/2306.10108} {2306.10108} \BibitemShut {NoStop}%
\bibitem [{\citenamefont {Shalgar}\ and\ \citenamefont
  {Tamborra}(2023{\natexlab{a}})}]{shalgar2023neutrino}%
  \BibitemOpen
  \bibfield  {author} {\bibinfo {author} {\bibfnamefont {S.}~\bibnamefont
  {Shalgar}}\ and\ \bibinfo {author} {\bibfnamefont {I.}~\bibnamefont
  {Tamborra}},\ }\href {\doibase 10.1103/PhysRevD.108.043006} {\bibfield
  {journal} {\bibinfo  {journal} {Phys. Rev. D}\ }\textbf {\bibinfo {volume}
  {108}},\ \bibinfo {pages} {043006} (\bibinfo {year} {2023}{\natexlab{a}})},\
  \Eprint {http://arxiv.org/abs/2206.00676} {2206.00676} \BibitemShut {NoStop}%
\bibitem [{\citenamefont {Shalgar}\ and\ \citenamefont
  {Tamborra}(2023{\natexlab{b}})}]{shalgar2023neutrino1}%
  \BibitemOpen
  \bibfield  {author} {\bibinfo {author} {\bibfnamefont {S.}~\bibnamefont
  {Shalgar}}\ and\ \bibinfo {author} {\bibfnamefont {I.}~\bibnamefont
  {Tamborra}},\ }\href {\doibase 10.1103/PhysRevD.107.063025} {\bibfield
  {journal} {\bibinfo  {journal} {Phys. Rev. D}\ }\textbf {\bibinfo {volume}
  {107}},\ \bibinfo {pages} {063025} (\bibinfo {year} {2023}{\natexlab{b}})},\
  \Eprint {http://arxiv.org/abs/2207.04058} {2207.04058} \BibitemShut {NoStop}%
\bibitem [{\citenamefont {Cornelius}\ \emph
  {et~al.}(2024{\natexlab{b}})\citenamefont {Cornelius}, \citenamefont
  {Shalgar},\ and\ \citenamefont {Tamborra}}]{cornelius2024neutrino}%
  \BibitemOpen
  \bibfield  {author} {\bibinfo {author} {\bibfnamefont {M.}~\bibnamefont
  {Cornelius}}, \bibinfo {author} {\bibfnamefont {S.}~\bibnamefont {Shalgar}},
  \ and\ \bibinfo {author} {\bibfnamefont {I.}~\bibnamefont {Tamborra}},\
  }\href {\doibase 10.1088/1475-7516/2024/11/060} {\bibfield  {journal}
  {\bibinfo  {journal} {JCAP}\ }\textbf {\bibinfo {volume} {11}},\ \bibinfo
  {pages} {060} (\bibinfo {year} {2024}{\natexlab{b}})},\ \Eprint
  {http://arxiv.org/abs/2407.04769} {arXiv:2407.04769 [astro-ph.HE]}
  \BibitemShut {NoStop}%
\bibitem [{\citenamefont {Johns}(2023)}]{johns2023thermodynamics}%
  \BibitemOpen
  \bibfield  {author} {\bibinfo {author} {\bibfnamefont {L.}~\bibnamefont
  {Johns}},\ }\href@noop {} {\  (\bibinfo {year} {2023})},\ \Eprint
  {http://arxiv.org/abs/2306.14982} {2306.14982} \BibitemShut {NoStop}%
\bibitem [{\citenamefont {Nagakura}\ \emph {et~al.}(2024)\citenamefont
  {Nagakura}, \citenamefont {Johns},\ and\ \citenamefont
  {Zaizen}}]{nagakura2024bgk}%
  \BibitemOpen
  \bibfield  {author} {\bibinfo {author} {\bibfnamefont {H.}~\bibnamefont
  {Nagakura}}, \bibinfo {author} {\bibfnamefont {L.}~\bibnamefont {Johns}}, \
  and\ \bibinfo {author} {\bibfnamefont {M.}~\bibnamefont {Zaizen}},\ }\href
  {\doibase 10.1103/PhysRevD.109.083013} {\bibfield  {journal} {\bibinfo
  {journal} {Phys. Rev. D}\ }\textbf {\bibinfo {volume} {109}},\ \bibinfo
  {pages} {083013} (\bibinfo {year} {2024})},\ \Eprint
  {http://arxiv.org/abs/2312.16285} {2312.16285} \BibitemShut {NoStop}%
\bibitem [{\citenamefont {Johns}(2024)}]{johns2024subgrid}%
  \BibitemOpen
  \bibfield  {author} {\bibinfo {author} {\bibfnamefont {L.}~\bibnamefont
  {Johns}},\ }\href@noop {} {\  (\bibinfo {year} {2024})},\ \Eprint
  {http://arxiv.org/abs/2401.15247} {2401.15247} \BibitemShut {NoStop}%
\bibitem [{\citenamefont {Johns}\ and\ \citenamefont
  {Nagakura}(2021)}]{johns2021fast}%
  \BibitemOpen
  \bibfield  {author} {\bibinfo {author} {\bibfnamefont {L.}~\bibnamefont
  {Johns}}\ and\ \bibinfo {author} {\bibfnamefont {H.}~\bibnamefont
  {Nagakura}},\ }\href {\doibase 10.1103/PhysRevD.103.123012} {\bibfield
  {journal} {\bibinfo  {journal} {Phys. Rev. D}\ }\textbf {\bibinfo {volume}
  {103}},\ \bibinfo {pages} {123012} (\bibinfo {year} {2021})},\ \Eprint
  {http://arxiv.org/abs/2104.04106} {2104.04106} \BibitemShut {NoStop}%
\bibitem [{\citenamefont {Nagakura}\ and\ \citenamefont
  {Johns}(2021)}]{nagakura2021new}%
  \BibitemOpen
  \bibfield  {author} {\bibinfo {author} {\bibfnamefont {H.}~\bibnamefont
  {Nagakura}}\ and\ \bibinfo {author} {\bibfnamefont {L.}~\bibnamefont
  {Johns}},\ }\href {\doibase 10.1103/PhysRevD.104.063014} {\bibfield
  {journal} {\bibinfo  {journal} {Phys. Rev. D}\ }\textbf {\bibinfo {volume}
  {104}},\ \bibinfo {pages} {063014} (\bibinfo {year} {2021})},\ \Eprint
  {http://arxiv.org/abs/2106.02650} {2106.02650} \BibitemShut {NoStop}%
\bibitem [{\citenamefont {Abbar}(2023)}]{abbar2023applications}%
  \BibitemOpen
  \bibfield  {author} {\bibinfo {author} {\bibfnamefont {S.}~\bibnamefont
  {Abbar}},\ }\href {\doibase 10.1103/PhysRevD.107.103006} {\bibfield
  {journal} {\bibinfo  {journal} {Phys. Rev. D}\ }\textbf {\bibinfo {volume}
  {107}},\ \bibinfo {pages} {103006} (\bibinfo {year} {2023})},\ \Eprint
  {http://arxiv.org/abs/2303.05560} {arXiv:2303.05560 [astro-ph.HE]}
  \BibitemShut {NoStop}%
\bibitem [{\citenamefont {Abbar}\ and\ \citenamefont
  {Nagakura}(2024)}]{abbar2024detecting}%
  \BibitemOpen
  \bibfield  {author} {\bibinfo {author} {\bibfnamefont {S.}~\bibnamefont
  {Abbar}}\ and\ \bibinfo {author} {\bibfnamefont {H.}~\bibnamefont
  {Nagakura}},\ }\href {\doibase 10.1103/PhysRevD.109.023033} {\bibfield
  {journal} {\bibinfo  {journal} {Phys. Rev. D}\ }\textbf {\bibinfo {volume}
  {109}},\ \bibinfo {pages} {023033} (\bibinfo {year} {2024})},\ \Eprint
  {http://arxiv.org/abs/2310.03807} {2310.03807} \BibitemShut {NoStop}%
\bibitem [{\citenamefont {Abbar}\ \emph
  {et~al.}(2024{\natexlab{a}})\citenamefont {Abbar}, \citenamefont {Wu},\ and\
  \citenamefont {Xiong}}]{abbar2024physicsinformed}%
  \BibitemOpen
  \bibfield  {author} {\bibinfo {author} {\bibfnamefont {S.}~\bibnamefont
  {Abbar}}, \bibinfo {author} {\bibfnamefont {M.-R.}\ \bibnamefont {Wu}}, \
  and\ \bibinfo {author} {\bibfnamefont {Z.}~\bibnamefont {Xiong}},\ }\href
  {\doibase 10.1103/PhysRevD.109.043024} {\bibfield  {journal} {\bibinfo
  {journal} {Phys. Rev. D}\ }\textbf {\bibinfo {volume} {109}},\ \bibinfo
  {pages} {043024} (\bibinfo {year} {2024}{\natexlab{a}})},\ \Eprint
  {http://arxiv.org/abs/2311.15656} {2311.15656} \BibitemShut {NoStop}%
\bibitem [{\citenamefont {Abbar}\ \emph
  {et~al.}(2024{\natexlab{b}})\citenamefont {Abbar}, \citenamefont {Wu},\ and\
  \citenamefont {Xiong}}]{abbar2024application}%
  \BibitemOpen
  \bibfield  {author} {\bibinfo {author} {\bibfnamefont {S.}~\bibnamefont
  {Abbar}}, \bibinfo {author} {\bibfnamefont {M.-R.}\ \bibnamefont {Wu}}, \
  and\ \bibinfo {author} {\bibfnamefont {Z.}~\bibnamefont {Xiong}},\ }\href
  {\doibase 10.1103/PhysRevD.109.083019} {\bibfield  {journal} {\bibinfo
  {journal} {Phys. Rev. D}\ }\textbf {\bibinfo {volume} {109}},\ \bibinfo
  {pages} {083019} (\bibinfo {year} {2024}{\natexlab{b}})},\ \Eprint
  {http://arxiv.org/abs/2401.17424} {2401.17424} \BibitemShut {NoStop}%
\bibitem [{\citenamefont {Froustey}\ \emph {et~al.}(2024)\citenamefont
  {Froustey}, \citenamefont {Richers}, \citenamefont {Grohs}, \citenamefont
  {Flynn}, \citenamefont {Foucart}, \citenamefont {Kneller},\ and\
  \citenamefont {McLaughlin}}]{froustey2024neutrino}%
  \BibitemOpen
  \bibfield  {author} {\bibinfo {author} {\bibfnamefont {J.}~\bibnamefont
  {Froustey}}, \bibinfo {author} {\bibfnamefont {S.}~\bibnamefont {Richers}},
  \bibinfo {author} {\bibfnamefont {E.}~\bibnamefont {Grohs}}, \bibinfo
  {author} {\bibfnamefont {S.}~\bibnamefont {Flynn}}, \bibinfo {author}
  {\bibfnamefont {F.}~\bibnamefont {Foucart}}, \bibinfo {author} {\bibfnamefont
  {J.~P.}\ \bibnamefont {Kneller}}, \ and\ \bibinfo {author} {\bibfnamefont
  {G.~C.}\ \bibnamefont {McLaughlin}},\ }\href {\doibase
  10.1103/PhysRevD.109.043046} {\bibfield  {journal} {\bibinfo  {journal}
  {Phys. Rev. D}\ }\textbf {\bibinfo {volume} {109}},\ \bibinfo {pages}
  {043046} (\bibinfo {year} {2024})},\ \Eprint
  {http://arxiv.org/abs/2311.11968} {2311.11968} \BibitemShut {NoStop}%
\bibitem [{\citenamefont {Grohs}\ \emph {et~al.}(2024)\citenamefont {Grohs},
  \citenamefont {Richers}, \citenamefont {Couch}, \citenamefont {Foucart},
  \citenamefont {Froustey}, \citenamefont {Kneller},\ and\ \citenamefont
  {McLaughlin}}]{grohs2024twomoment}%
  \BibitemOpen
  \bibfield  {author} {\bibinfo {author} {\bibfnamefont {E.}~\bibnamefont
  {Grohs}}, \bibinfo {author} {\bibfnamefont {S.}~\bibnamefont {Richers}},
  \bibinfo {author} {\bibfnamefont {S.~M.}\ \bibnamefont {Couch}}, \bibinfo
  {author} {\bibfnamefont {F.}~\bibnamefont {Foucart}}, \bibinfo {author}
  {\bibfnamefont {J.}~\bibnamefont {Froustey}}, \bibinfo {author}
  {\bibfnamefont {J.}~\bibnamefont {Kneller}}, \ and\ \bibinfo {author}
  {\bibfnamefont {G.}~\bibnamefont {McLaughlin}},\ }\href {\doibase
  10.3847/1538-4357/ad13f2} {\bibfield  {journal} {\bibinfo  {journal}
  {Astrophys. J.}\ }\textbf {\bibinfo {volume} {963}},\ \bibinfo {pages} {11}
  (\bibinfo {year} {2024})},\ \Eprint {http://arxiv.org/abs/2309.00972}
  {2309.00972} \BibitemShut {NoStop}%
\bibitem [{\citenamefont {Richers}\ \emph
  {et~al.}(2021{\natexlab{b}})\citenamefont {Richers}, \citenamefont
  {Willcox},\ and\ \citenamefont {Ford}}]{richers2021neutrino}%
  \BibitemOpen
  \bibfield  {author} {\bibinfo {author} {\bibfnamefont {S.}~\bibnamefont
  {Richers}}, \bibinfo {author} {\bibfnamefont {D.}~\bibnamefont {Willcox}}, \
  and\ \bibinfo {author} {\bibfnamefont {N.}~\bibnamefont {Ford}},\ }\href
  {\doibase 10.1103/PhysRevD.104.103023} {\bibfield  {journal} {\bibinfo
  {journal} {Phys. Rev. D}\ }\textbf {\bibinfo {volume} {104}},\ \bibinfo
  {pages} {103023} (\bibinfo {year} {2021}{\natexlab{b}})},\ \Eprint
  {http://arxiv.org/abs/2109.08631} {2109.08631} \BibitemShut {NoStop}%
\bibitem [{\citenamefont {Richers}\ \emph {et~al.}(2024)\citenamefont
  {Richers}, \citenamefont {Froustey}, \citenamefont {Ghosh}, \citenamefont
  {Foucart},\ and\ \citenamefont {Gomez}}]{richers2024asymptoticstate}%
  \BibitemOpen
  \bibfield  {author} {\bibinfo {author} {\bibfnamefont {S.}~\bibnamefont
  {Richers}}, \bibinfo {author} {\bibfnamefont {J.}~\bibnamefont {Froustey}},
  \bibinfo {author} {\bibfnamefont {S.}~\bibnamefont {Ghosh}}, \bibinfo
  {author} {\bibfnamefont {F.}~\bibnamefont {Foucart}}, \ and\ \bibinfo
  {author} {\bibfnamefont {J.}~\bibnamefont {Gomez}},\ }\href {\doibase
  10.1103/PhysRevD.110.103019} {\bibfield  {journal} {\bibinfo  {journal}
  {Phys. Rev. D}\ }\textbf {\bibinfo {volume} {110}},\ \bibinfo {pages}
  {103019} (\bibinfo {year} {2024})},\ \Eprint
  {http://arxiv.org/abs/2409.04405} {arXiv:2409.04405 [astro-ph.HE]}
  \BibitemShut {NoStop}%
\bibitem [{\citenamefont {George}\ \emph {et~al.}(2023)\citenamefont {George},
  \citenamefont {Lin}, \citenamefont {Wu}, \citenamefont {Liu},\ and\
  \citenamefont {Xiong}}]{george2023cose}%
  \BibitemOpen
  \bibfield  {author} {\bibinfo {author} {\bibfnamefont {M.}~\bibnamefont
  {George}}, \bibinfo {author} {\bibfnamefont {C.-Y.}\ \bibnamefont {Lin}},
  \bibinfo {author} {\bibfnamefont {M.-R.}\ \bibnamefont {Wu}}, \bibinfo
  {author} {\bibfnamefont {T.~G.}\ \bibnamefont {Liu}}, \ and\ \bibinfo
  {author} {\bibfnamefont {Z.}~\bibnamefont {Xiong}},\ }\href {\doibase
  10.1016/j.cpc.2022.108588} {\bibfield  {journal} {\bibinfo  {journal}
  {Comput. Phys. Commun.}\ }\textbf {\bibinfo {volume} {283}},\ \bibinfo
  {pages} {108588} (\bibinfo {year} {2023})},\ \Eprint
  {http://arxiv.org/abs/2203.12866} {2203.12866} \BibitemShut {NoStop}%
\bibitem [{\citenamefont {Banerjee}\ \emph {et~al.}(2011)\citenamefont
  {Banerjee}, \citenamefont {Dighe},\ and\ \citenamefont
  {Raffelt}}]{banerjee2011linearized}%
  \BibitemOpen
  \bibfield  {author} {\bibinfo {author} {\bibfnamefont {A.}~\bibnamefont
  {Banerjee}}, \bibinfo {author} {\bibfnamefont {A.}~\bibnamefont {Dighe}}, \
  and\ \bibinfo {author} {\bibfnamefont {G.}~\bibnamefont {Raffelt}},\ }\href
  {\doibase 10.1103/PhysRevD.84.053013} {\bibfield  {journal} {\bibinfo
  {journal} {Phys. Rev. D}\ }\textbf {\bibinfo {volume} {84}},\ \bibinfo
  {pages} {053013} (\bibinfo {year} {2011})},\ \Eprint
  {http://arxiv.org/abs/1107.2308} {1107.2308} \BibitemShut {NoStop}%
\bibitem [{\citenamefont {Raffelt}\ \emph {et~al.}(2013)\citenamefont
  {Raffelt}, \citenamefont {Sarikas},\ and\ \citenamefont
  {Seixas}}]{raffelt2013axial}%
  \BibitemOpen
  \bibfield  {author} {\bibinfo {author} {\bibfnamefont {G.}~\bibnamefont
  {Raffelt}}, \bibinfo {author} {\bibfnamefont {S.}~\bibnamefont {Sarikas}}, \
  and\ \bibinfo {author} {\bibfnamefont {D.~D.~S.}\ \bibnamefont {Seixas}},\
  }\href {\doibase 10.1103/PhysRevLett.111.091101} {\bibfield  {journal}
  {\bibinfo  {journal} {Phys. Rev. Lett.}\ }\textbf {\bibinfo {volume} {111}},\
  \bibinfo {pages} {091101} (\bibinfo {year} {2013})},\ \Eprint
  {http://arxiv.org/abs/1305.7140} {1305.7140} \BibitemShut {NoStop}%
\bibitem [{\citenamefont {Hunter}(2007)}]{matplotlib}%
  \BibitemOpen
  \bibfield  {author} {\bibinfo {author} {\bibfnamefont {J.~D.}\ \bibnamefont
  {Hunter}},\ }\href {\doibase 10.1109/MCSE.2007.55} {\bibfield  {journal}
  {\bibinfo  {journal} {Computing in Science \& Engineering}\ }\textbf
  {\bibinfo {volume} {9}},\ \bibinfo {pages} {90} (\bibinfo {year}
  {2007})}\BibitemShut {NoStop}%
\bibitem [{\citenamefont {van~der Walt}\ \emph {et~al.}(2011)\citenamefont
  {van~der Walt}, \citenamefont {Colbert},\ and\ \citenamefont
  {Varoquaux}}]{numpy}%
  \BibitemOpen
  \bibfield  {author} {\bibinfo {author} {\bibfnamefont {S.}~\bibnamefont
  {van~der Walt}}, \bibinfo {author} {\bibfnamefont {S.~C.}\ \bibnamefont
  {Colbert}}, \ and\ \bibinfo {author} {\bibfnamefont {G.}~\bibnamefont
  {Varoquaux}},\ }\href {\doibase 10.1109/MCSE.2011.37} {\bibfield  {journal}
  {\bibinfo  {journal} {Computing in Science \& Engineering}\ }\textbf
  {\bibinfo {volume} {13}},\ \bibinfo {pages} {22} (\bibinfo {year}
  {2011})}\BibitemShut {NoStop}%
\bibitem [{\citenamefont {Turk}\ \emph {et~al.}(2011)\citenamefont {Turk},
  \citenamefont {Smith}, \citenamefont {Oishi}, \citenamefont {Skory},
  \citenamefont {Skillman}, \citenamefont {Abel},\ and\ \citenamefont
  {Norman}}]{turk2010yt}%
  \BibitemOpen
  \bibfield  {author} {\bibinfo {author} {\bibfnamefont {M.~J.}\ \bibnamefont
  {Turk}}, \bibinfo {author} {\bibfnamefont {B.~D.}\ \bibnamefont {Smith}},
  \bibinfo {author} {\bibfnamefont {J.~S.}\ \bibnamefont {Oishi}}, \bibinfo
  {author} {\bibfnamefont {S.}~\bibnamefont {Skory}}, \bibinfo {author}
  {\bibfnamefont {S.~W.}\ \bibnamefont {Skillman}}, \bibinfo {author}
  {\bibfnamefont {T.}~\bibnamefont {Abel}}, \ and\ \bibinfo {author}
  {\bibfnamefont {M.~L.}\ \bibnamefont {Norman}},\ }\href {\doibase
  10.1088/0067-0049/192/1/9} {\bibfield  {journal} {\bibinfo  {journal}
  {Astrophys. J. Suppl.}\ }\textbf {\bibinfo {volume} {192}},\ \bibinfo {pages}
  {9} (\bibinfo {year} {2011})},\ \Eprint {http://arxiv.org/abs/1011.3514}
  {arXiv:1011.3514 [astro-ph.IM]} \BibitemShut {NoStop}%
\end{thebibliography}
%

\appendix
\section{Angular distributions of asymptotic states}
\begin{figure*}[t]
\centering
\includegraphics[width=.49\textwidth]{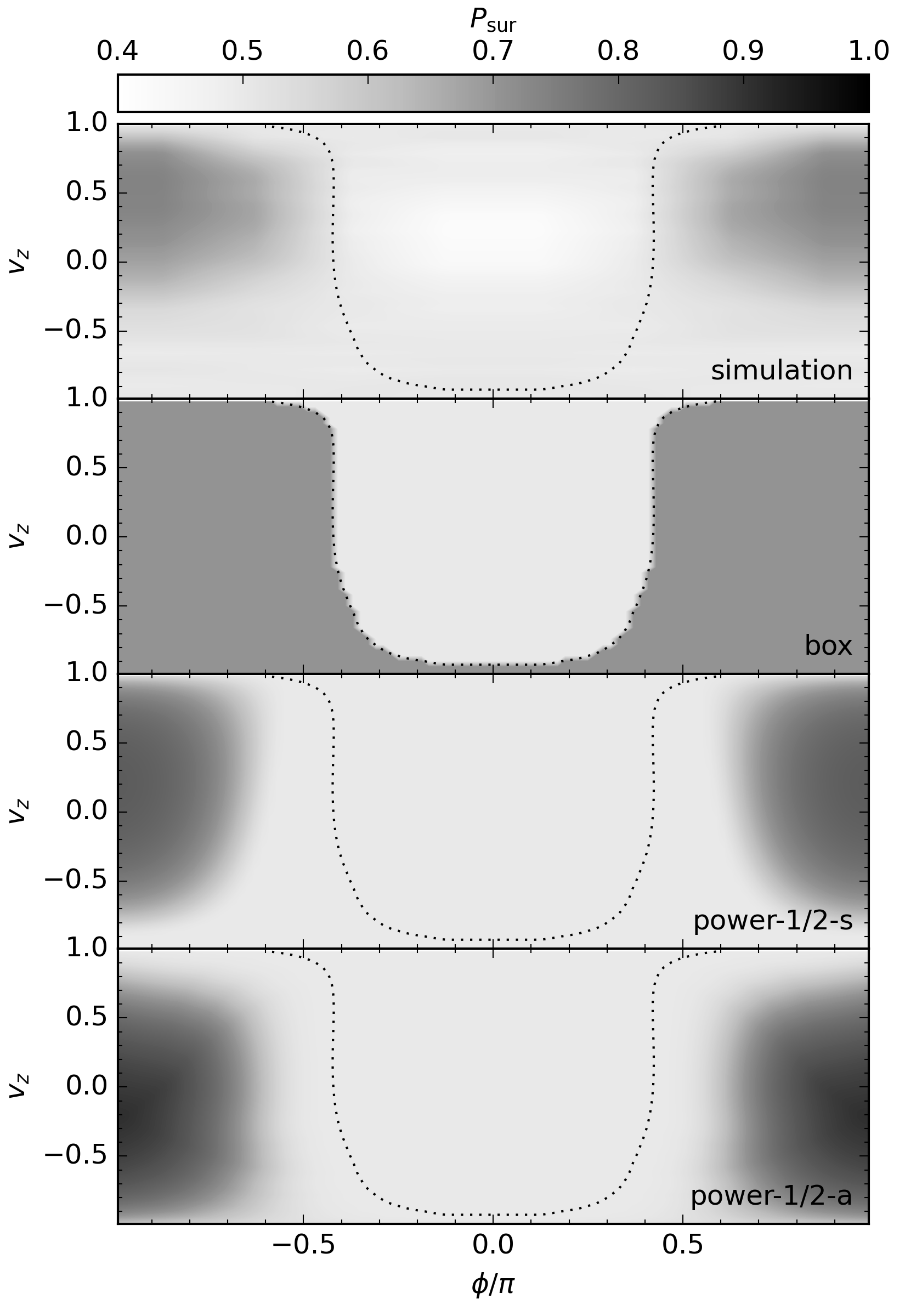}
\includegraphics[width=.49\textwidth]{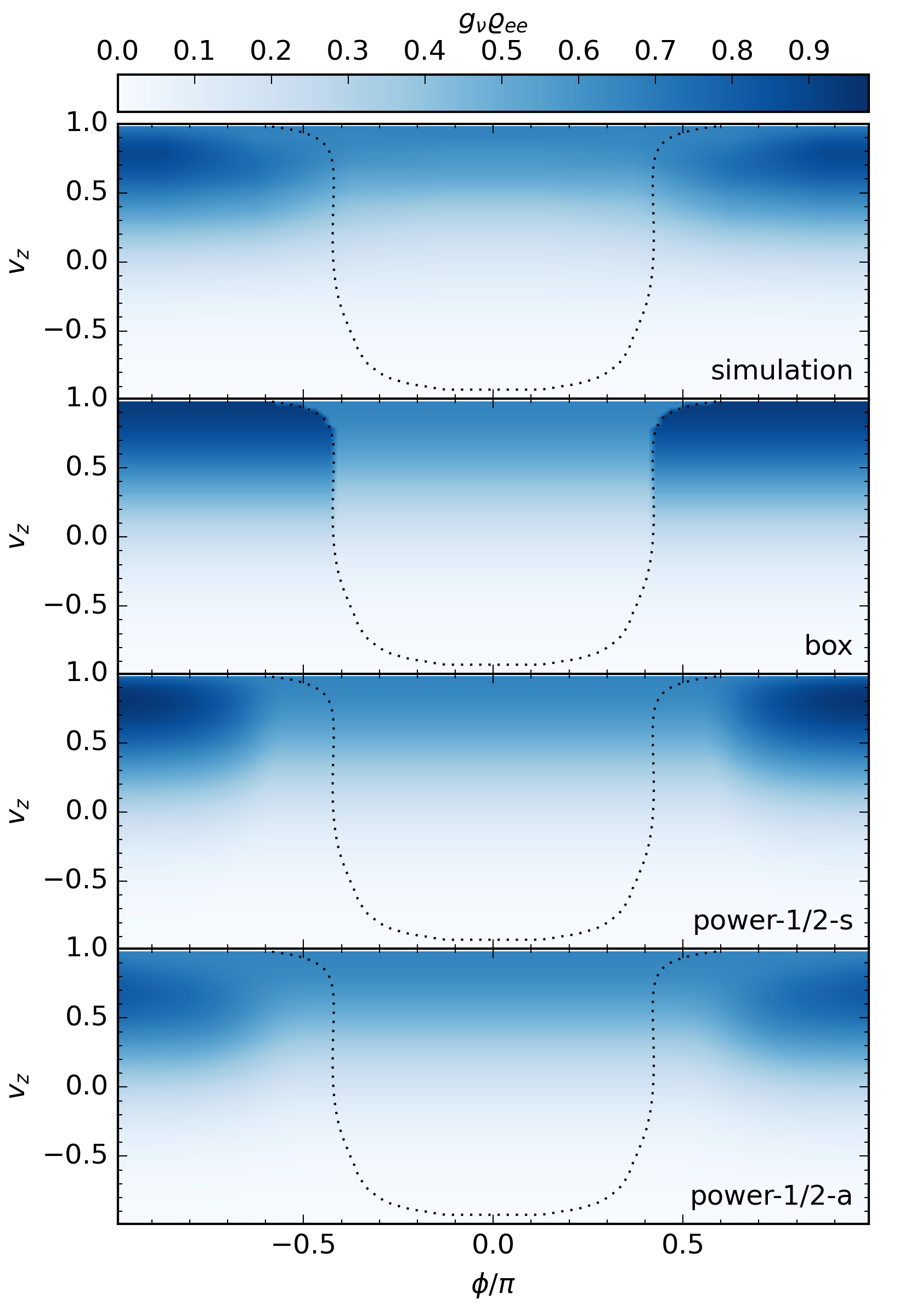}
\caption{
Comparison for the angular distributions of the survival probability $P_\mathrm{sur}$ (left panels) and $g_\nu \varrho_{ee}$ (right panels) in the coarse-grained asymptotic states from the simulations with those using the box-like, power-1/2-s, and power-1/2-a prescriptions for $\theta_r=30^\circ$.
Black dotted curves denote the initial ELN angular crossing.
\label{fig:asymptotic_dist}}
\end{figure*}

In addition to the angular moments, we compare the full angular distributions in the asymptotic states given by different analytical prescriptions to the coarse-grained ones from simulation for the case of $\theta_r=30^\circ$ in Fig.~\ref{fig:asymptotic_dist}.
The left column compares the survival probability $P_{\rm sur}$ obtained from spatially averaged $\langle P_{\rm sur}\rangle_V$ to $P^b_{\rm sur}$ [Eq.~\eqref{eq:box_presc}], $P_{\rm sur}^{p-s}$ [Eq.~\eqref{eq:p12_presc}],  and $P_{\rm sur}^{p-a}$ [Eq.~\eqref{eq:p12a_presc}], labeled by ``simulation'', ``box'', ``power-1/2-s'', and ``power-1/2-a'', respectively. 
The right column shows the corresponding electron neutrino angular distributions, $g_\nu P_{\rm sur}$, for all cases.
For better illustration, these distributions are interpolated onto finer velocity grids.
Black dotted curves denote the initial ELN angular crossing for all panels.

As described in the main text, the angular distribution of the survival probability obtained from simulation shows 
that near flavor equilibration ($P_\mathrm{sur}\sim 0.5$) is achieved in one of the angular domains separated by the ELN crossing contour, with a slight flavor overconversion around $\phi\sim 0$ and $v_z\sim 0$. 
In the other domain, $P_{\rm sur}$ increases smoothly from $\sim 0.5$ from the ELN crossing contour to larger values of $\sim 0.8$ at $\phi\sim \pi$ and $\theta\sim 0.5$.
For $P_{\rm sur}$ given by different prescriptions, both the power-1/2-s and power-1/2-a schemes qualitatively capture the smooth transition obtained in simulation, although both schemes predict fewer flavor conversions around $\phi\sim 0$ and $v_z\lesssim 0$.
As for the box-like scheme, $P_{\rm sur}$ takes two distinct values, which transit abruptly at the crossing contour by construction.

For the $\nu_e$ angular distribution, $g_\nu P_{\rm sur}$, since $g_\nu$ is more forward peaked in $v_z$, the values of $g_\nu P_{\rm sur}$ are suppressed in $v_z\lesssim 0$ in all cases as shown in the right column of Fig.~\ref{fig:asymptotic_dist}. 
However,    
the above abrupt transition in the box-like scheme remains clearly visible in $g_\nu P_{\rm sur}$, which results in larger deviation of the predicted angular moments discussed in the main text. 
For the two power-1/2 prescriptions, the overall shapes of $g_\nu \varrho_{ee}$ agree better with the simulation results than that from the box prescription, leading to smaller deviations in angular moments discussed in the main text.

\end{document}